\documentclass[12pt]{article}
\usepackage{epsf,amsfonts,amssymb,epsfig,amsmath,graphics,slashed}
\usepackage{hyperref,graphicx, subfig,hep}
\usepackage{rotating}
\usepackage{color}
\def\be{\begin{equation}}
\def\ee{\end{equation}}
\def\bea{\begin{eqnarray}}
\def\eea{\end{eqnarray}}

\newcommand{\nn}{\notag \\}

\begin{document}

\makeatletter
\renewcommand{\theequation}{\thesection.\arabic{equation}}
\@addtoreset{equation}{section}
\makeatother

\baselineskip 18pt

\begin{titlepage}

\vfill

\begin{flushright}
Imperial/TP/2012/JG/05\\
MIT-CTP 4421\\
\end{flushright}

\vfill

\begin{center}
   \baselineskip=16pt
   {\Large\bf  Competing orders in M-theory: \\
     \vskip .2cm
     superfluids, stripes and metamagnetism}
  \vskip 1.5cm
      Aristomenis Donos$^1$, Jerome P. Gauntlett$^1$, Julian Sonner$^2$ and Benjamin Withers$^3$\\
   \vskip .6cm
      \begin{small}
      \textit{$^1$Blackett Laboratory, 
        Imperial College\\ London, SW7 2AZ, U.K.}
        \end{small}\\*[.6cm]
   \begin{small}
      \textit{$^2$C.T.P., Massachusetts Institute of Technology\\ Cambridge, MA 02139, U.S.A.}
        \end{small}\\*[.6cm]
     \textit{$^3$Centre for Particle Theory and Department of Mathematical Sciences\\ University of Durham, South Road, Durham, DH1 3LE, U.K.}


\end{center}

\vfill

\begin{center}
\textbf{Abstract}
\end{center}

\begin{quote}
We analyse the infinite class of $d=3$ CFTs dual to skew-whiffed $AdS_4\times SE_7$ solutions of $D=11$ supergravity
at finite temperature and charge density and in the presence of a magnetic field. 
We construct black hole solutions corresponding to
the unbroken phase, and at zero temperature some of these
become dyonic domain walls of an Einstein-Maxwell-pseudo-scalar theory interpolating between $AdS_4$ in the UV and new families of dyonic $AdS_2\times\mathbb{R}^2$ solutions in the IR. 
The black holes exhibit both diamagnetic and paramagnetic behaviour.
We analyse superfluid and striped instabilities and show that for large enough values of the magnetic field the
superfluid instability disappears while the striped instability remains. For larger values of the magnetic field there is also a first-order metamagnetic phase transition and at zero temperature these black hole solutions exhibit
hyperscaling violation in the IR with dynamical exponent $z=3/2$ and $\theta=-2$.
\end{quote}

\vfill

\end{titlepage}

\setcounter{equation}{0}


\section{Introduction}
The AdS/CFT correspondence allows us to study strongly coupled quantum field theories
at finite temperature by studying appropriate black hole solutions of a dual gravitational theory. 
One focus has been the study of conformal field theories with AdS duals when held at finite chemical potential with
respect to a global $U(1)$ symmetry.  Depending on the details of the gravitational theory, whose matter content includes
a $U(1)$ gauge-field, various types of novel phases are possible, corresponding to the existence of fascinating new classes
of electrically charged black branes. 

One well-studied possibility is that the CFT undergoes a superfluid phase transition. The first constructions were
in the context of a bottom-up, Einstein-Maxwell theory of gravity coupled to a charged scalar field \cite{Gubser:2008px,Hartnoll:2008vx,Hartnoll:2008kx}
and then extended to top-down constructions in \cite{Gauntlett:2009dn,Gubser:2009qm}.
Another possibility is that the CFT undergoes a phase transition to a spatially modulated phase, in which translation invariance is spontaneously broken.
This has been discussed in the context of electrically charged black holes \cite{Nakamura:2009tf,Ooguri:2010kt,Donos:2011bh,Donos:2011ff} 
and also magnetically charged black holes \cite{Donos:2011qt,Donos:2011pn}. Other work, utilising the brane probe approximation, can be found in 
\cite{Domokos:2007kt,Ooguri:2010xs,Bayona:2011ab,Bergman:2011rf}. 
In some special examples in $D=5$ it is possible to construct fully back reacted black hole solutions 
that are spatially modulated, with a helical structure, by solving ODEs \cite{Donos:2012gg,Donos:2012wi} (see also \cite{Iizuka:2012iv}). 
However, for most cases, including the $D=4$ examples of interest in this paper, one will need to solve PDEs which is technically 
more challenging\footnote{The PDEs associated with some of
the holographic striped black holes discovered in \cite{Donos:2011bh} were recently studied in \cite{Rozali:2012es}.}.

A simple diagnostic for the existence of new branches of 
black hole solutions at finite temperature can often
be obtained by analysing the zero temperature limit of the unbroken, normal phase black hole solutions.
Indeed, this is possible when the ground state that is approached in this limit has a finite entropy density, $s\ne 0$.
For example, focussing on $D=4$, if the zero temperature limit is described by a domain wall solution that interpolates between $AdS_4$ in the UV and $AdS_2\times \mathbb{R}^2$ in the IR,
as in the AdS-RN black brane solution,
one can look for modes of the $AdS_2\times \mathbb{R}^2$ solution that violate the $AdS_2$ BF bound. If such modes exist, the $T=0$ domain wall solution will also be unstable and hence, by continuity, so will the finite-temperature black hole solutions, for low enough temperatures. 
To determine the critical temperature at which this instability sets in and where a new branch of black holes appears, one should look for static, normalisable, linearised perturbations about the finite temperature unbroken phase black hole solutions. While this approach is conceptually straightforward, one can encounter situations, as we will here, where the critical temperature is so low that it
is hard to stabilise the numerical integration to find its precise value. One can nevertheless take the conceptual point of view that new phases emerge at low temperatures due to the destabilising effect of the finite ground-state entropy at $T=0$ and thus that the qualitative features of the finite-temperature phase structure are dictated by the nature of the mode spectrum in the $s\neq 0$ ground state itself. One is thus naturally led to investigate the competition of possible sources of instabilities in the finite-entropy state.

In this paper we will employ these methods to analyse the competition between superfluid and spatially modulated or ``striped"\footnote{The word ``striped" arises because the linearised mode for the spatially modulated branch has a striped structure, being translationally invariant along a given direction. 
At the linearised level one can also superpose such modes in different directions, losing this translation invariance. 
In order to determine the precise nature of the spatially modulated
phase will require constructing the backreacted black holes. }
phases \cite{Donos:2011bh} in a top-down setting. More specifically we will consider a $D=4$ model in which both of these instabilities are present, and then switch
on a magnetic field, aiming to suppress the superfluid instability while maintaining the striped instability. This scenario is in fact realised
and we also find, for very large magnetic fields, a first-order metamagnetic phase transition 
at non-zero temperature. Such transitions, involving a discontinuous jump
in the magnetisation of the system and not usually associated with symmetry breaking, are seen in a variety of materials, such as heavy fermion systems.
They have also been discussed in a holographic context involving probe branes in \cite{Lifschytz:2009sz}, but as far as
we are aware our construction with $T\ne 0$ is the first in a gravitational setting. 
Following these solutions down to zero temperature, we find that they exhibit hyperscaling violation in the far IR.
Holographic hyperscaling violating solutions have been studied in
\cite{Charmousis:2010zz,Ogawa:2011bz,Huijse:2011ef} and top-down constructions appear\footnote{In fact the 
$T\to 0$ limit of the uncharged and also the charged normal phase black holes of \cite{Gauntlett:2009bh} approach
top-down hyperscaling violating solutions in the IR with $z=1$ and $\theta=-1$.}
 in
\cite{Dong:2012se,Singh:2012un,Narayan:2012hk,Dey:2012tg}.
Our solutions become purely magnetic
in the IR with dynamical exponent $z=3/2$ and hyperscaling violation exponent $\theta=-2$. In particular,
the entropy scales like $s\propto T^{8/3}$ as $T\to 0$. 

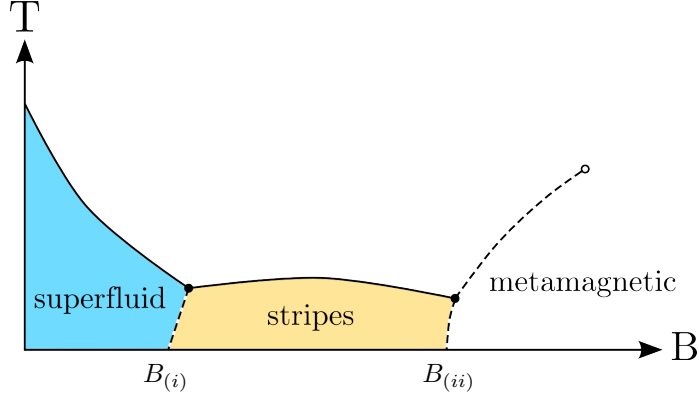
\begin{figure}[h!]
\begin{center}
\begin{picture}(0.1,0.25)(0,0)
\put(63,-7){\makebox(0,0){{\footnotesize$B_{(i)}$}}}
\put(170,-7){\makebox(0,0){{\footnotesize $B_{(ii)}$}}}
\end{picture}
\includegraphics[width=0.6\textwidth]{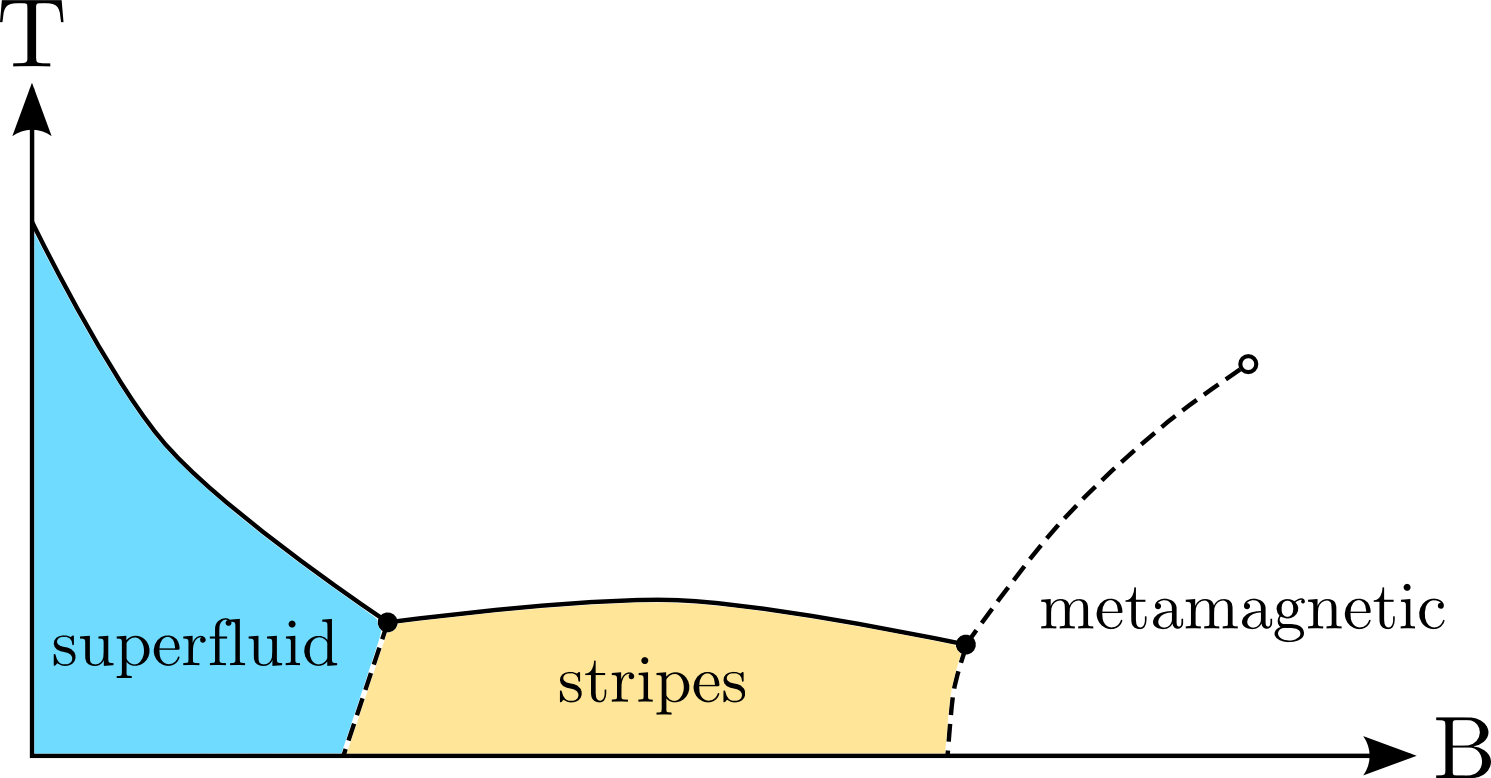}
\vskip 1em
\caption{A schematic figure of a plausible phase diagram as a function of applied magnetic field. The solid lines denote second order phase
transitions and the dashed lines first order. The two solid dots denote tri-critical points and the open circle a second order critical point where
the metamagnetic transition ends.
The phase diagram is symmetric under $B\to -B$.
For $B>B_{(ii)}$ as $T\to 0$ the solutions exhibit hyperscaling violation in the IR 
with $z=3/2$ and $\theta=-2$.
  \label{TBphase}}
\end{center}
\end{figure}

A schematic picture of the likely phase diagram
incorporating our findings is given in figure \ref{TBphase}; there are some assumptions going into this figure which will be discussed in detail in the text.
Notice that the metamagnetic transition ends in a critical point which is second-order (as in the liquid-vapour case) or higher, at finite temperature.

The $D=4$ top-down model that we shall consider 
couples the metric with a gauge field, a charged scalar field and a neutral pseudo-scalar field, $\sigma$ and
arises from a consistent Kaluza-Klein (KK) reduction 
on an arbitrary $D=7$ Sasaki-Einstein space $SE_7$ \cite{Gauntlett:2009bh,Gauntlett:2009zw}. This means that 
any solution of this $D=4$ theory can be uplifted
to $D=11$ on an arbitrary $SE_7$ space leading to an infinite class of $D=11$ solutions. In particular,
the vacuum $AdS_4$ solution uplifts to the skew-whiffed $AdS_4\times SE_7$ solutions which should\footnote{A dual CFT will exist provided that the
skew-whiffed $AdS_4\times SE_7$ solution is stable. It is known that skew-whiffed solutions are perturbatively stable \cite{Duff:1984sv}. A discussion of
possible dual CFTs can be found in \cite{Forcella:2011pp}.}
be dual to $d=3$ CFTs with an abelian global symmetry and, generically, no supersymmetry,
apart from the special case when $SE_7=S^7$ which preserves all of the supersymmetry.

The unbroken phase of these $d=3$ CFTs at finite temperature, $T$, and chemical potential with respect to the abelian symmetry, $\mu$, is described
by the electrically charged AdS-RN black brane solution. A branch of superfluid black hole solutions appears at 
a critical temperature \cite{Denef:2009tp}
and the fully back-reacted solutions were constructed in \cite{Gauntlett:2009bh}. On the other hand at very low temperatures the AdS-RN black brane 
also connects with a spatially modulated ``striped"
branch of black holes \cite{Donos:2011bh}. This was demonstrated by constructing BF violating modes in the $AdS_2\times\mathbb{R}^2$ solution arising in the IR limit of the $T=0$ AdS-RN black hole solution. By analysing linearised modes about the finite temperature AdS-RN black holes, it was shown in 
\cite{Donos:2011bh} that the critical temperature for this transition is extremely low, but the precise value was not found. With these results we cannot be certain about what happens to the 
superfluid phase as it is cooled. The simplest possibility is that the system stays on the superfluid branch of black holes all the way down to zero temperature. However, it is possible that the system moves to a striped phase at low temperatures 
either via a first order or a second order phase transition. It is also possible that there are transitions to other phases. To establish this
one would need to know, in principle, all of the black holes that exist at low temperatures, including the fully back reacted spatially
modulated black holes. In this paper, we will not address these issues as they remain technically out of reach.

Here, instead, we will analyse the class of $d=3$ skew-whiffed CFTS at finite $T,\mu$ after switching on a magnetic field, $B$. 
The presence of both electric and magnetic fields provides a source for the $D=4$ neutral, pseudo-scalar field, $\sigma$,
and this leads to a very rich structure for the unbroken, normal-phase black hole solutions (i.e. with vanishing charged scalar field). 
Indeed the study of these dyonic black hole solutions and related domain wall solutions that appear at zero temperature, both 
of which are solutions  
of a simple Einstein-Maxwell-pseudo-scalar theory in $D=4$, and which are of interest in their own right,
will be the focus
for much of our analysis\footnote{It is worth comparing and contrasting our charged pseudo-scalar black holes
with the charged, dilatonic black holes 
i.e. with neutral scalar fields, which have been studied in a holographic context in
many works, including \cite{Taylor:2008tg,Gubser:2009qt,Goldstein:2009cv,Charmousis:2010zz,Goldstein:2010aw}.}.

We begin by showing that the $D=4$ Einstein-Maxwell-pseudo-scalar model admits two 
families of dyonic $AdS_2\times\mathbb{R}^2$ solutions carrying, generically, both electric and magnetic charges and $\sigma\ne 0$. One family, the ``electric family", contains the purely electrically charged $AdS_2\times\mathbb{R}^2$ solution, arising in the electric AdS-RN solution at $T=0$, 
while the other, the ``magnetic family", contains the purely magnetic solution, arising in the magnetic AdS-RN solution at $T=0$.
We then investigate which of these solutions can arise as the IR limit of a domain wall solution that asymptotes to 
$AdS_4$ in the UV, with deformation data given by $\mu, B$. Such domain walls are
possible zero temperature limits of finite temperature black holes describing the unbroken phase. 
Interestingly
we find that there can be more than one domain wall solution with the same UV data $\mu, B$ and yet
different $AdS_2\times\mathbb{R}^2$ solutions in the IR. These solutions differ in the normalisable data in the UV and the parameters
governing the irrelevant operators in the IR.
Another surprising result is that while we find that some of the domain wall solutions can be heated up to arbitrarily high temperatures some cannot.
Further insight into the unbroken phase can be obtained by calculating the free energy of the black holes and,
in particular, we will see the first order metamagnetic phase transition appearing for large $B$. In contrast to the dyonic 
AdS-RN black holes of Einstein-Maxwell theory which are always strongly diamagnetic \cite{Hartnoll:2008kx,Denef:2009yy}
(as we review in appendix A), 
we will also see paramagnetic behaviour.

The plan of the rest of the paper is as follows. In section \ref{sec2} we introduce the top-down Einstein-Maxwell pseudo-scalar model
that will be used in sections \ref{sec2} - \ref{dybhs}.
The new dyonic-pseudo-scalar $AdS_2\times \mathbb{R}^2$ solutions are presented in section \ref{axdysol}.
In section \ref{anandtherm} we introduce our ansatz for domain wall and black hole solutions and discuss some aspects of
the thermodynamics. In particular we include a discussion of the definition of magnetisation and magnetic susceptibility.
Sections \ref{ddws} and \ref{dybhs} construct domain wall and black hole solutions, including some discussion of the magnetisation properties
of the domain walls and the metamagnetic phase transition. 
Section \ref{sasi} analyses the striped and superfluid instabilities of the $AdS_2\times \mathbb{R}^2$ and the black hole
solutions. 
In analysing the striped instabilities, which do not involve the charged scalar fields, we find the surprising result 
that, after a scaling of the wave-number, the spectrum of perturbations is the same for all dyonic $AdS_2\times \mathbb{R}^2$ solutions.
At the end of section \ref{sasi} we summarise our conclusions about the full phase diagram, which lead to 
figure \ref{TBphase}. We briefly conclude in section \ref{fcomms}.
Finally, we have one appendix where we calculate the magnetisation and susceptibility of dyonic black holes in Einstein-Maxwell theory
to compare with the results that we obtain in our model.

\section{Top down Einstein-Maxwell-pseudo-scalar model}\label{sec2}
For most of the paper we will consider the $D=4$ model of \cite{Gauntlett:2009bh}
which couples the metric to a gauge-field, $A$, and a neutral pseudo-scalar field, $\sigma$. The
action is given by 
\begin{align}\label{action}
S= &\frac{1}{16\pi G} \int d^4 x \sqrt{-g} \left(R - \frac{1}{2}(\partial \sigma)^2  - \frac{\tau(\sigma)}{4} F^2 - V(\sigma)\right)
+\frac{1}{32\pi G}\int \vartheta(\sigma) F\wedge F 
\end{align}
where $F=dA$ and
\begin{align}\label{fnsub}
V(\sigma)  \equiv   -24\cosh{\frac{\sigma}{\sqrt{3}} }\,,\qquad
\tau(\sigma)  \equiv \frac{1}{{\cosh\sqrt{3}\sigma}}\,,\qquad
\vartheta(\sigma)  \equiv  \tanh\sqrt{3}\sigma\,.
\end{align}
The inclusion of the charged field of \cite{Gauntlett:2009bh} will be treated later when we discuss superfluid instabilities. Note that to 
compare\footnote{To compare with \cite{Donos:2011bh} we should set $\sigma=\sqrt{2}\varphi$, rescale the metric $g^{here}=(1/2) g^{there}$, identify the potentials $V^{here}=4V^{there}$ and also $\vartheta^{here}=-\vartheta^{there}$.} with 
\cite{Gauntlett:2009bh}  we should set $h=\tanh{\frac{\sigma}{\sqrt{3}}}$ and also rescale the gauge field $A^{here}=2 A^{there}$.
Occasionally we will find it convenient to use the field $h$ instead of $\sigma$ in some plots.

 The equations of motion are given by
\begin{align}\label{eom}
&R_{ab}=\tfrac{1}{2}\partial_a\sigma\partial_b\sigma
+\frac{V}{2} g_{ab}
+\frac{\tau}{2}\left(F^2_{ab}-\tfrac{1}{4}g_{ab}F^2\right),\nn
&d\left(\tau*F\right)=d\vartheta\wedge F,\nn
&d*d\sigma+V'*1+\frac{\tau'}{2}F\wedge *F-\frac{\vartheta}{2} 'F\wedge F= 0\,.
\end{align}
Any solution of these equations of motion, for the specific functions given in \eqref{fnsub}, can be uplifted on an arbitrary $SE_7$ manifold to obtain
an exact solution of $D=11$ supergravity using the formulae in \cite{Gauntlett:2009bh}.
For example the basic vacuum $AdS_4$ solution with $\sigma=A=0$ and radius squared $1/4$ uplifts to
the standard skew-whiffed $AdS_4\times SE_7$ solution, which doesn't preserve any supersymmetry 
except in the special case of $SE_7=S^7$ in which it preserves all of the supersymmetry. 

It is interesting to observe that given\footnote{More generally, we require
that $(\tau,V)$ and $\vartheta$ are even and odd functions of $\sigma$, respectively, that are analytic at $\sigma=0$.} \eqref{fnsub}, the equations of motion exhibit
the following electric/magnetic duality 
symmetry
\begin{align}\label{dualtrans}
F\to-\tau(\sigma)*F+\vartheta(\sigma)F,\qquad 
\sigma\to -\sigma
\end{align}
The origin of this symmetry should be related to the fact that the action comes from a truncation of an $N=2$ supergravity theory 
\cite{Gauntlett:2009zw,Gauntlett:2009bh}.
We also observe that, for the specific functions given in \eqref{fnsub}, it is only consistent to set $\sigma=0$ for configurations
that have $F\wedge F=0$. For such configurations, the equations of motion collapse to those of Einstein-Maxwell theory.
We will concentrate on the functions given in \eqref{fnsub} but many of our results have simple generalisations for different choices of functions.

\section{Dyonic $AdS_2\times \mathbb{R}^2$ solutions}\label{axdysol}
We consider the following ansatz
\def\Ea{E}
\begin{align}\label{ads2sols}
ds^2&=L^2ds^2\left(AdS_2\right)+ ds^2(\mathbb{R}^2)\,,\nn
F&=-\Ea L^2\mathrm{Vol}(AdS_2)+B\mathrm{Vol}(\mathbb{R}^2)\,,
\end{align}
where $ds^2\left(AdS_2\right)$ and $\mathrm{Vol}(AdS_2)$ are the metric and volume form for a unit radius $AdS_2$, and 
$\sigma,L,E$ and $B$ are constants, and the minus sign appearing is for later convenience.
Substituting into the equations of motion for \eqref{actansatz} we are led to the following algebraic conditions 
\begin{align}\label{con1}
E^2+B^2=-\frac{2V}{\tau},\qquad\qquad
\frac{\tau'}{2}(E^2-B^2)-\vartheta' \Ea B- V'=0\,,
\end{align}
with the $AdS_2$ radius given by 
\begin{align}\label{con2}
L^{-2}=-{V}\,.
\end{align}
Notice that these equations are invariant under simultaneously flipping the sign of $B$ and $\sigma$. In addition
the duality transformation \eqref{dualtrans} corresponds to
\begin{align}
\Ea\to \tau B +\vartheta \Ea,\qquad B\to -\tau \Ea+\vartheta B,\qquad \sigma\to -\sigma
\end{align}

Up to flipping the sign of both $E$ and $B$, for the specific functions given in \eqref{fnsub}, these relations define two one-parameter families of dyonic $AdS_2\times \mathbb{R}^2$ solutions, labelled by $\sigma$, which we have summarised in figure 2.
Notice that one family, which we call the ``electric family", contains a purely electric solution, while the other  
``magnetic family", contains a purely magnetic solution. As we discuss below, 
the purely electric and magnetic $AdS_2\times \mathbb{R}^2$ solutions, both of which have $\sigma=0$, arise as
the near horizon limit of the standard electric and magnetic AdS-RN black brane solutions, respectively. 

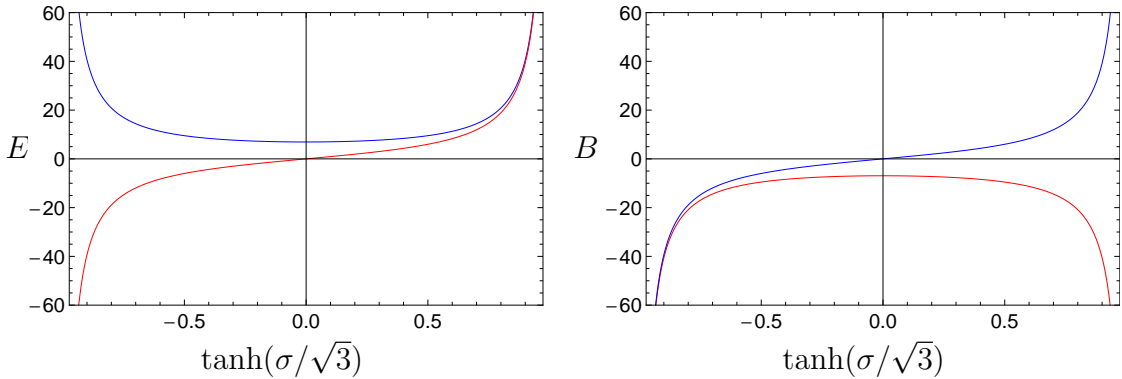
\begin{figure}[h!]
\begin{center}
\begin{picture}(0.1,0.25)(0,0)
\put(100,-10){\makebox(0,0){{\rm tanh}($\sigma/\sqrt{3}$)}}
\put(0,70){\makebox(0,0){$E$}}
\put(320,-10){\makebox(0,0){{\rm tanh}($\sigma/\sqrt{3}$)}}
\put(215,70){\makebox(0,0){$B$}}
\end{picture}
\includegraphics[width=0.45\textwidth]{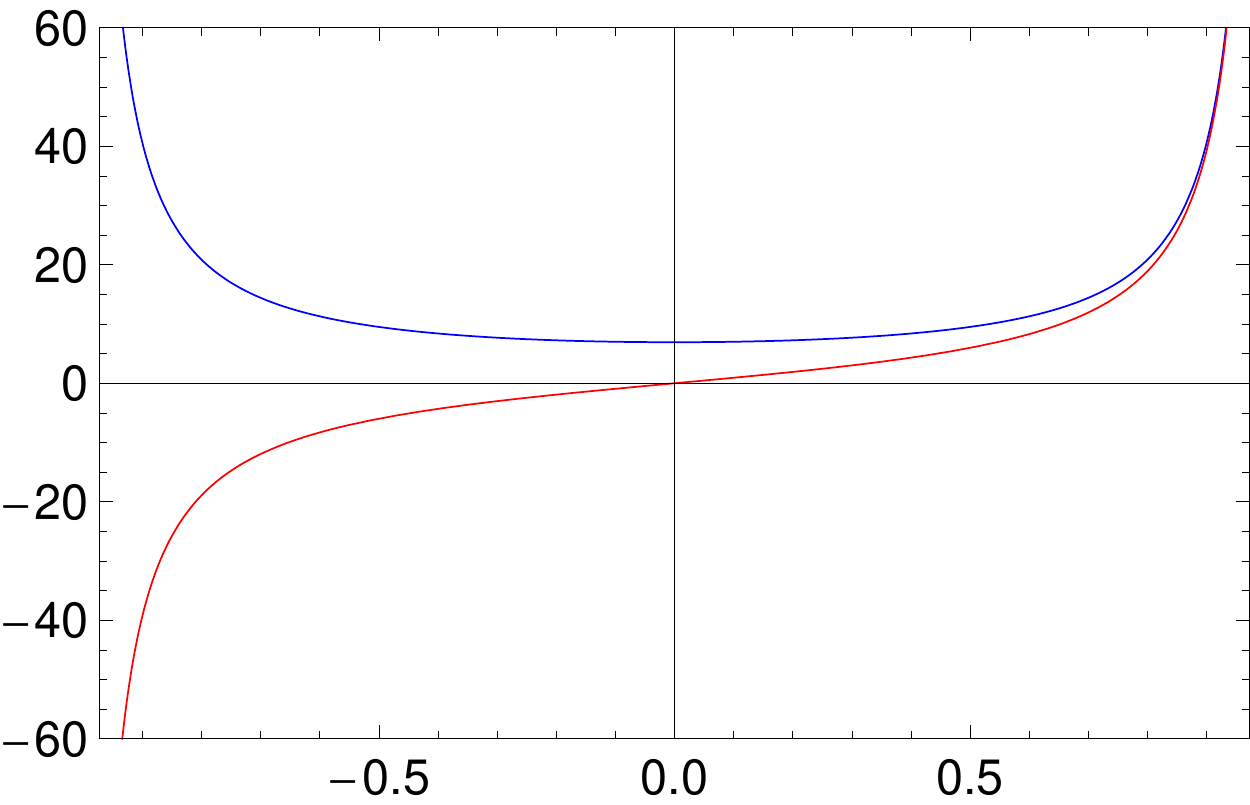}\hskip2em
\includegraphics[width=0.45
\textwidth]{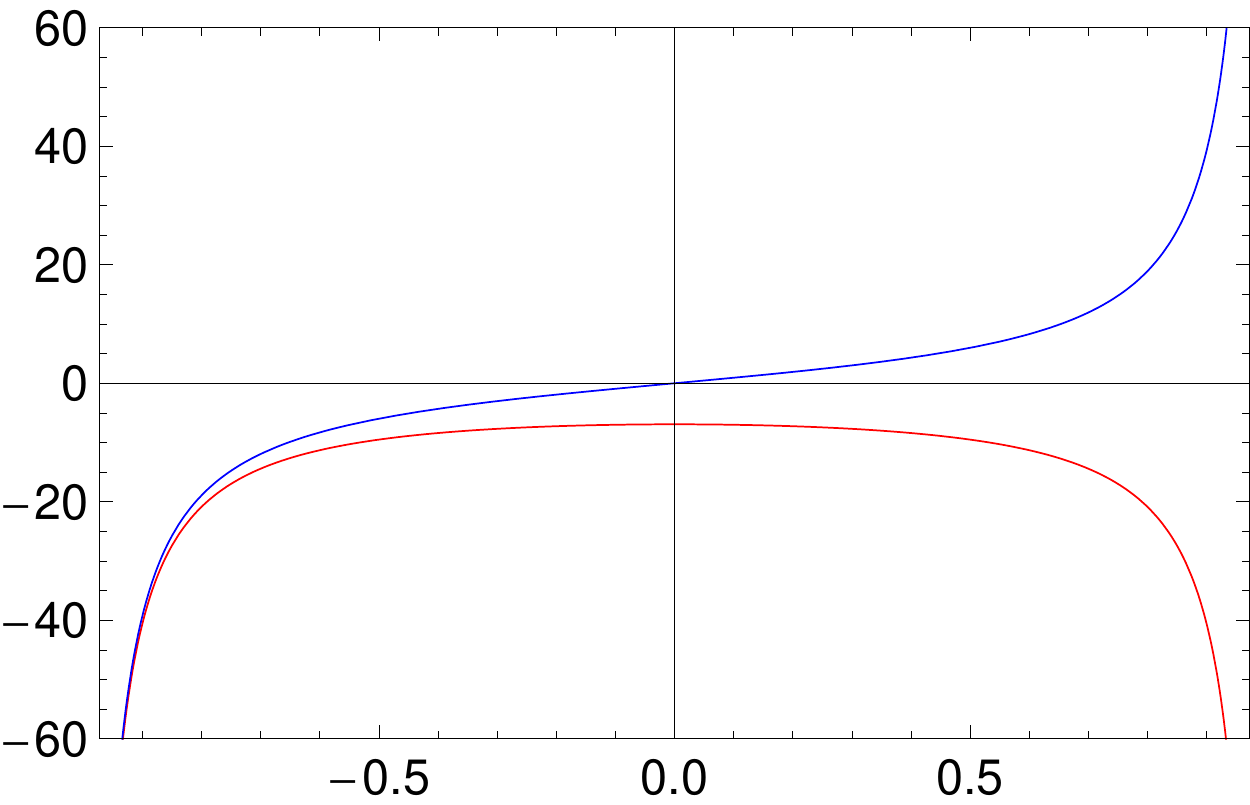}
\vskip1em
\caption{Two families of dyonic $AdS_2\times \mathbb{R}^2$ solutions. The left and right panels  
display the dependence of $E$ and $B$ on the value of the pseudo-scalar $\sigma$, respectively. For convenience of presentation we use
$\tanh(\sigma/\sqrt{3})$ for the horizontal axis. The blue line is the electric family which contains a purely electric solution
and the red line is the magnetic family which contains a purely magnetic solution. There are two more families of solutions obtained by simultaneously
flipping the signs of $E,B$.
  \label{fam}}
\end{center}
\end{figure}

\section{Ansatz and thermodynamics}\label{anandtherm}

\subsection{Ansatz for domain wall and black hole solutions}
In the sequel we will consider the following ansatz
\begin{align}\label{solans}
ds^2 &= - e^{-\beta(r)}g(r) dt^2 + \frac{dr^2}{g(r)} + r^2 (dx^2+dy^2)\,,\nn
A &= \phi(r) dt + B \tfrac{1}{2}(x dy-ydx)\,,\nn
\sigma&=\sigma(r)\,.
\end{align}
After substituting into the equations of motion \eqref{eom} we obtain differential equations for the functions
$\beta,g,\phi$ and $\sigma$. These equations can also be obtained by substituting the ansatz directly into the action \eqref{action},
leading to 
\begin{align}\label{actansatz}
S&=c_0\int dr r^2e^{-\beta/2}\Big[-g''+g'\left(\frac{3}{2}\beta'-\frac{4}{r}\right)+g\left(\beta''-\frac{1}{2}(\beta')^2+2\frac{\beta'}{r}-\frac{2}{r^2}\right)\nn
&
\qquad\qquad-\frac{1}{2}g\sigma'^2 +\frac{1}{2}\tau(\sigma)\left(e^{\beta}\phi'^2-\frac{B^2}{r^4}\right)-V(\sigma)
-B\vartheta(\sigma)\frac{e^{\beta/2}\phi'}{r^2}\Big]\,,
\end{align}
where $c_0=(16\pi G)^{-1}\int dt dx dy$, and then varying.
It is helpful to observe that the ansatz and hence the equations of motion are invariant under the two scalings
\begin{align}\label{scsym}
&t\to\lambda t,\qquad e^{\beta}\to \lambda^2e^{\beta},\qquad\phi\to \lambda^{-1}\phi\,;\nn
&r\to\lambda r,\quad (t,x,y)\to \lambda^{-1}(t,x,y),\quad g\to\lambda^2 g,\quad \phi\to\lambda \phi,\quad B\to \lambda^2 B\,.
\end{align}
We also note the symmetry
\begin{align}\label{bsym}
B\to -B,\qquad \sigma\to -\sigma
\end{align}

When $\sigma=0$, the standard electrically charged AdS-RN black brane solves the equations of motion. It is given by 
\begin{align}\label{eladsrn}
g=4r^2-(4r_+^2+\frac{\mu^2}{4})\frac{r_+}{r}+\frac{\mu^2 r_+^2}{4 r^2},\qquad \phi=\mu(1-\frac{r_+}{r})
\end{align}
with $\beta =B=0$. When $\mu=4\sqrt{3}r_+$, the temperature $T=0$ and  
the solution is a domain wall interpolating between $AdS_4$ in the UV and the purely electrically charged $AdS_2\times\mathbb{R}^2$ solution in the IR given
in \eqref{ads2sols},\eqref{con1},\eqref{con2} with $L^2=1/24$ and $E=-4\sqrt{3}$.
Although we will be principally interested in black hole solutions with non-vanishing electric charge, we recall here 
the purely magnetic AdS-RN black brane solution:
\begin{align}\label{magadsrn}
g=4r^2-(4r_+^2+\frac{B^2}{4r_+^2})\frac{r_+}{r}+\frac{B^2}{4r^2},
\end{align}
with $\beta =\phi=0$.
The $T=0$ limit is when $B=4\sqrt{3}r_+^2$ and the solution becomes a domain wall
approaching the purely magnetically charged $AdS_2\times\mathbb{R}^2$ solution in the IR given
in \eqref{ads2sols},\eqref{con1},\eqref{con2} with $L^2=1/24$ and $B=4\sqrt{3}$.

Notice that the standard dyonic AdS-RN black hole is not a solution to the equations of motion, since the presence of both electric and magnetic fields
sources the pseudo-scalar $\sigma$. We will construct dyonic black hole and domain wall solutions with $\sigma\ne 0$ numerically in later sections.

\subsection{Asymptotic expansions}
\subsubsection{UV expansion}
We are interested in studying the $d=3$ CFTs with chemical potential $\mu$ and magnetic field $B$. In the UV, as $r\to\infty$,
we will impose the following expansions
\begin{align}\label{expan}
g &= 4r^2 +  \sigma_1^2 - \frac{1}{2r}\left(\varepsilon - 4\sigma_1\sigma_2\right)+\ldots\,,\nn
\beta &= \beta_a+\frac{\sigma_1^2}{4r^2}+\frac{2\sigma_1\sigma_2}{3r^3}\ldots\,,\nn
\phi &= e^{-\frac{\beta_a}{2}}\left(\mu - \frac{q}{r} -\frac{\sqrt{3}B\sigma_1}{2r^2}+\ldots\right)\,,\nn
\sigma &= \frac{\sigma_1}{r}+\frac{\sigma_2}{r^2}+\frac{5\sigma_1^3}{72r^3}\ldots\,.
\end{align}
Note that for simplicity\footnote{For the special case of uplifting the $D=4$ solutions on an $S^7$, the choice
$\Delta({\cal O}_\sigma)=2$ is associated with maximal supersymmetry.}
we focus on the quantisation of the pseudo-scalar so that the dual operator has dimension
$\Delta({\cal O}_\sigma)=2$. For the most part, we will consider the CFT with no deformation in the UV by ${\cal O}_\sigma$
corresponding to setting $\sigma_1=0$. For simplicity of presentation, we have set 
\begin{align}
16\pi G=1
\end{align}
The appropriate factors can easily be reinstated if required. It is
also worth repeating here that the radius of the asymptotic $AdS_4$ is 1/2.

\subsubsection{IR expansion: Finite Temperature}
In this case we are interested in black hole solutions with regular event horizons located at $r=r_+$ where there is an analytic expansion of the 
form
\begin{align}\label{bhexp}
g&=g_+(r-r_+)+\dots\,,\nn
\beta&=\beta_++\dots\,,\nn
\phi&=\phi_+(r-r_+)+\dots\,,\nn
\sigma&=\sigma_++\dots\,.
\end{align}
This expansion is fixed by 4 constants, $\beta_+,\phi_+,\sigma_+$ and $r_+$ with, for example,
\begin{align}
g_+=12 r_+\cosh\frac{\sigma_+}{\sqrt 3}-\frac{B^2+r_+e^{\beta_+}\phi_+^2}{4\cosh\sqrt{3}\sigma_+}\,.
\end{align}
A black hole solution is then specified by 6 UV parameters (with $\sigma_1=0$) and 4 IR parameters.
From \eqref{actansatz} we have two first order equations of motion, for $g$, $\beta$, and two second order equations,
for $\phi,\sigma$ and so a solution is specified by 6 integration constants. 
Taking into account the scaling symmetries \eqref{scsym}, we expect two-parameter families of black hole solutions which
can be labelled by temperature $T$ and magnetic field $B$, or better, by the dimensionless
quantities $T/\mu$ and $B/\mu^2$.

For later use we notice that the equation of motion for $\phi$ arising from \eqref{actansatz} (i.e. the 
equation of motion for the gauge-field in \eqref{eom}) can be integrated from $r=r_+$ to $r=\infty$ and from \eqref{bhexp} and
\eqref{expan}
we obtain the charge conservation condition 
\begin{align}\label{chgecon}
q=\tau(\sigma_+) e^{\beta_+/2}\phi_+ r_+^2-B\vartheta(\sigma_+)\,,
\end{align}
with the first and second terms 
on the right hand side arising from the $F^2$ term and the $F\wedge F$ terms in
the action \eqref{action}, respectively. 
This condition can be satisfied in the far IR of a black hole solution in
the zero temperature limit in different ways. 
For example, we will construct dyonic black hole solutions which approach domain walls
at $T=0$ with non-zero entropy density, which will get
get contributions from both terms. We will also construct black hole solutions with vanishing entropy density
as $T\to 0$ with all of the contribution to $q$ coming from the $F\wedge F$ term.

\subsubsection{IR expansion: Zero Temperature}

We are also interested in domain wall solutions that asymptote in the IR, as $r\to r_+$, to the dyonic $AdS_2\times\mathbb{R}^2$ solutions
given in \eqref{ads2sols}-\eqref{con2}. We focus on the following expansion 
\begin{align}\label{dwexp}
g(r) &= g_0\, (r-r_+)^2 +\ldots\nn
\sigma(r)&= \sigma_0 + \ldots + \sigma_+\, (r-r_+)^{\Delta-1}+\ldots\nn
\phi(r) &= \phi_0\, (r-r_+) + \ldots + \sigma_+ \delta \phi \, (r-r_+)^{\Delta}+\ldots\nn
\beta(r) &= \beta_0 + \ldots + \sigma_+ \delta \beta\,  (r-r_+)^{\Delta-1}+\ldots\,.
\end{align}
where $g_0$ is determined via
\be
g_0 = - V(\sigma_0)\,,
\ee
together with the relations
\be\label{relations}
E^2 + \tilde B^2 = -\frac{2 V(\sigma_0)}{\tau(\sigma_0)},\quad \frac{\tau'(\sigma_0)}{2}(E^2-\tilde B^2)-\vartheta'(\sigma_0) \Ea\tilde B  - V'(\sigma_0)=0\,,
\ee
where $\Ea \equiv  e^{\frac{\beta_0}{2}}\phi_0$ and $\tilde B\equiv B/r_+^2$. Also,
$\sigma_+$ parametrises a deformation by an irrelevant operator of dimension $\Delta$, with
\be\label{dimirop}
\Delta =\frac{1}{6}(3+\sqrt{105})\,,
\ee
which we discuss further below.
As $r\to r_+$ this expansion approaches the exact dyonic $AdS_2\times\mathbb{R}^2$ solutions
given in \eqref{ads2sols}-\eqref{con2}. 
The irrelevant deformation with $\Delta$ as \eqref{dimirop} is obtained by linearising
about an exact dyonic $AdS_2\times\mathbb{R}^2$ solution. We find that the corresponding mode has
\bea
\delta \beta &=& -2\sqrt{3} \tanh{\frac{\sigma_0}{\sqrt{3}}}\,,\nn
\delta \phi &=& e^{\frac{-\beta_0}{2}}\frac{\sqrt{3}}{2\Delta}\frac{\tilde{B} +2E\sinh{\sqrt{3}\sigma_0}-2E\sinh{\frac{\sigma_0}{\sqrt{3}}}}{\cosh{\sqrt{3}\sigma_0}}\,.
\eea

A domain wall solution is specified by 6 UV parameters (with $\sigma_1=0$) and 3 IR parameters (i.e. $\sigma_0,\phi_0,\beta_0,r_+,\sigma_+$ subject
to the two constraints \eqref{relations}). Now from \eqref{actansatz} we have two first order equations of motion, for $g$, $\beta$, and two second order equations, for $\phi,\sigma$ and so a solution is specified by 6 integration constants. 
Taking into account the scaling symmetries \eqref{scsym}, we expect a one-parameter family of black hole solutions which
can be labelled by the magnetic field $B$.

\subsection{Thermodynamics}\label{thermsec}
Building on \cite{Hartnoll:2007ai,Hartnoll:2008kx}, 
we generalise the discussion of \cite{Gauntlett:2009bh} to include $B\ne 0$.
We analytically continue by setting $t= -i \tau$, together with $I = -i S$. We can then obtain two expressions for the
on-shell action for the class of solutions we are studying. The first expression is given by the integral of a total derivative
\be\label{firstexpr}
I_{OS}= {\Delta\tau vol_2}\int  dr  \left[r^2e^{-\frac{\beta}{2}}\left(g'-g\beta'-\tau(\sigma) e^\beta \phi \phi'\right) +B\vartheta(\sigma)  \phi\right]' \,.
\ee
The second can be written 
\begin{align}\label{secondexpr}
I_{OS}={\Delta\tau vol_2}\int  dr\left\{\left[  2r g e^{-\beta/2}   \right]' + B^2r^{-2}e^{-\beta/2}\tau(\sigma)  +B\vartheta(\sigma)\phi' \right\}\,,
\end{align}
which is an integral of a total derivative only in the special case that $B= 0$.

We define the total action $I_{Tot}$ via
\begin{align}
I_{Tot}=I+I_{ct}\,,
\end{align}
where the boundary counter term action is
\be
I_{ct} =  \int d\tau d^2x \sqrt{g_{\infty}}\left(-2K + 8 + \sigma^2\right)\,.
\ee
We next define the thermodynamic potential $W\equiv T [I_{tot}]_{OS}\equiv w vol_2$ where
the temperature of the black hole is given by
\begin{align}
T=\frac{e^{\beta_a/2}}{4\pi}[g'e^{-\beta/2}]_{r=r_+}\,.
\end{align}
Corresponding to the expression \eqref{firstexpr}, 
and using the expansion \eqref{expan} as well as \eqref{bhexp}, we obtain
\be
w= \varepsilon-\mu q-Ts\,.
\ee
On the other hand, corresponding to \eqref{secondexpr}
we find the alternative expression
\begin{align}\label{other}
w=-(\frac{\varepsilon}{2}+2\sigma_1\sigma_2)+{e^{\beta_a/2}}\int  dr\left\{B^2r^{-2}e^{-\beta/2}\tau(\sigma)  +B\vartheta(\sigma)\phi' \right\}\,.
\end{align}
The equality of these two expressions gives a Smarr type formula, which we shall return to below.

We next consider an on-shell variation of the total action, as in \cite{Gauntlett:2009bh}, and
deduce that $w=w(T,\mu,\sigma_1,B)$ with
\begin{align}
\delta w=-s\delta T-q\delta\mu-4\sigma_2\delta\sigma_1-m\delta B\,,
\end{align}
where the entropy density, $s$, is given by
\begin{align}
s={4\pi r_+^2}\,,
\end{align}
and the magnetisation per unit volume, $m\equiv -\partial w/\partial B$ at constant $T,\mu,\sigma_1$, is given by 
\begin{align}\label{mag}
m=-{e^{\beta_a/2}}\int  dr\left\{Br^{-2}e^{-\beta/2}\tau(\sigma)  +\vartheta(\sigma)\phi' \right\}\,.
\end{align}
Observe that the axionic-like coupling $\vartheta$ in \eqref{action} can give rise to magnetisation even when $B=0$.

The holographic stress tensor of \cite{Balasubramanian:1999re}, given by
\begin{align}
T_{i}{}^j=(2 r^3)[-2K_{i}{}^{j}+2\delta_{i}{}^j(2K-8-\sigma^2)]\,,
\end{align}
can be calculated and we find
\begin{align}\label{st}
T_t{}^t&=-\varepsilon\,,\nn
T_x{}^x=T_y{}^y&=\frac{\varepsilon}{2}+2\sigma_1\sigma_2\,.
\end{align}
Thus $\varepsilon$ is the energy of the system. We also see that when $\sigma_1=0$, corresponding to no deformation by ${\cal O}_\sigma$, but with
$B,\mu\ne0$, the stress tensor is traceless.
Using the general result that $w=-p$, where $p$ is the pressure, a comparison of \eqref{other}, \eqref{mag}
\eqref{st} reveals that the content of the Smarr relationship can be written 
\begin{align}
p&=\frac{\varepsilon}{2}+2\sigma_1\sigma_2+mB\,,\nn
&=T_x{}^x+mB\,,
\end{align}
similar to what was noted in \cite{Hartnoll:2008kx}.
The magnetic susceptibility, $\chi_m$, is defined by
$\chi_m=\partial m/\partial B=-\partial^2w/\partial B^2$ with the derivatives taken at constant $T,\mu,\sigma_1$.

\section{Dyonic Domain wall solutions}
\label{ddws}
In this section we construct domain wall solutions that 
approach $AdS_4$ in the UV as in \eqref{expan}, and we set $\sigma_1=0$
corresponding to no deformation of the CFT by the operator ${\cal O}_\sigma$. We also use the scaling \eqref{scsym} to set
$\beta_a=0$. In the IR they approach one of the dyonic
$AdS_2\times\mathbb{R}^2$ solutions discussed in section \ref{axdysol} via the expansion \eqref{dwexp}. As discussed above, 
we expect a one-parameter family of solutions which we label by the value of the dimensionless quantity
$B/\mu^2$. 
For definiteness we will choose $\mu>0$ and focus on $B\ge 0$. 
For $\sigma_1=0$, the solutions with $B\le 0$
can be obtained using
the symmetry \eqref{bsym}.

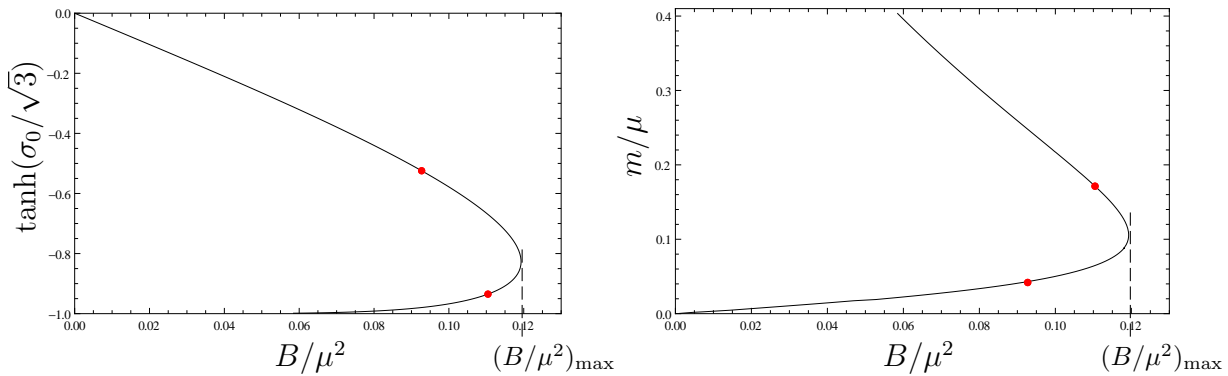
\begin{figure}[h!]
\begin{center}
\begin{picture}(0.1,0.25)(0,0)
\put(100,-10){\makebox(0,0){$B/\mu^2$}}
\put(192,-10){\makebox(0,0){{\footnotesize $(B/\mu^2)_{\rm max}$}}}
\multiput(180,-2)(0,7){5}{\line(0,1){5}}
\put(-8,70){\makebox(0,0){\begin{sideways}{\rm tanh}($\sigma_0/\sqrt{3}$)\end{sideways}}}
\end{picture}\includegraphics[width=0.45 \textwidth]{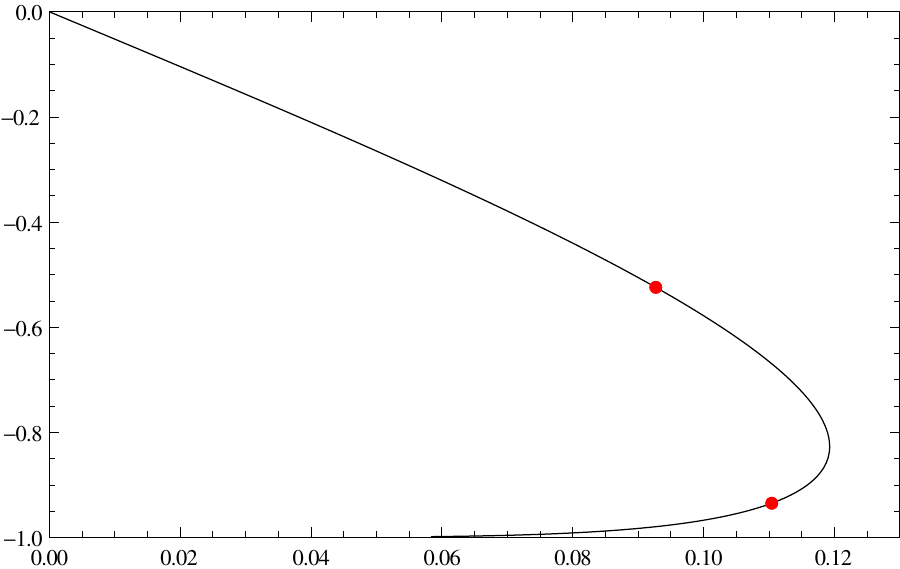}\qquad\quad
\begin{picture}(0.1,0.25)(0,0)
\put(100,-10){\makebox(0,0){$B/\mu^2$}}
\put(192,-10){\makebox(0,0){{\footnotesize $(B/\mu^2)_{\rm max}$}}}
\multiput(180,-2)(0,7){7}{\line(0,1){5}}
\put(-8,70){\makebox(0,0){\begin{sideways}$m/\mu$\end{sideways}}}
\end{picture}\includegraphics[width=0.45
\textwidth]{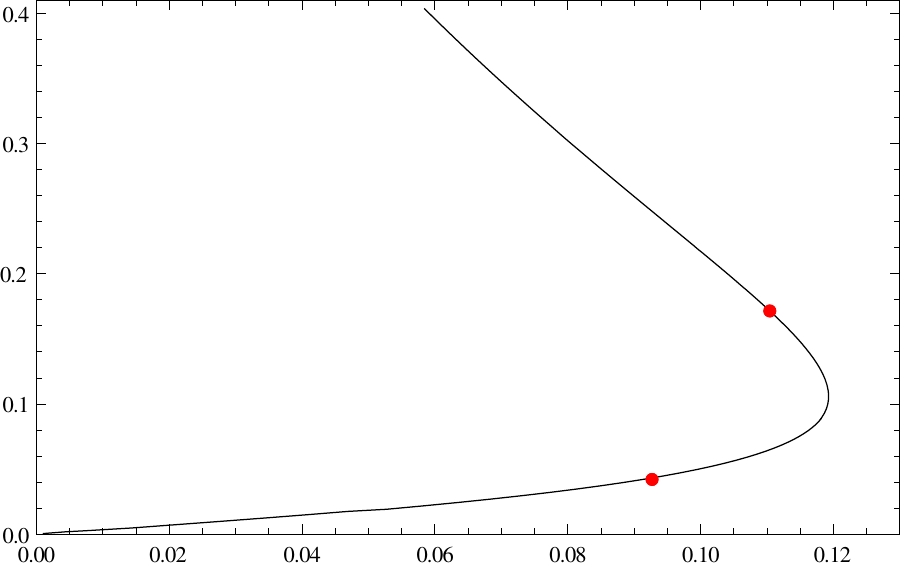}
\vskip1em
\caption{Left panel: the one parameter family of domain wall solutions interpolating between $AdS_4$ in the UV, with deformation
data $(\mu,B)$, and dyonic $AdS_2\times\mathbb{R}^2$ solutions in the electric family in the IR, labelled by $\sigma_0$. For convenience
of presentation the vertical axis is given by $\tanh(\sigma_0/\sqrt{3})$. There can be two domain wall solutions for given $(\mu,B)$ and
the upper branch has smaller free energy and is thermodynamically preferred. It is expected that the lower branch continues down to 
$B/\mu^2\to 0$. The red dots indicate superfluid instabilities, discussed in section \ref{secsfluid}, with the solutions being unstable to the left of the dots. Right panel: a plot of the magnetisation $m/\mu$ as a function of $B/\mu^2$. Observe that the magnetisation is always positive corresponding to paramagnetism.
  \label{dwallfig}}
\end{center}
\end{figure}

We first consider domain wall solutions that in the IR approach $AdS_2\times\mathbb{R}^2$ solutions lying in the electric family. The simplest
case is the purely electric $AdS_2\times\mathbb{R}^2$ solution for which the unique domain wall solution is simply the $T=0$ limit
of the electrically charged AdS-RN black brane \eqref{eladsrn}. As we switch on $B$ we find a one-parameter family of solutions as given
in figure \ref{dwallfig}. 
The first interesting feature is that the domain wall solutions only exist up to a maximum value of $B/\mu^2$ given by
$(B/\mu^2)_{\rm max}\approx 0.12$. The second interesting feature is that for fixed values of $B/\mu^2\ne 0$ there can be two distinct domain wall solutions.
These domain walls have the same UV deformation data $\mu,B$ but are distinguished in having different values of
$\varepsilon, q,\sigma_2$ in the UV expansion \eqref{expan}, 
corresponding to different values of the energy, charge and $\langle {\cal O}_\sigma\rangle$. After calculating the free
energy $w$ for these solutions we find that the upper branch in the left panel in figure \ref{dwallfig} is always thermodynamically preferred.
The numerical results suggest that this picture persists for all values of $B/\mu^2<(B/\mu^2)_{\rm{max}}$. 
Figure \ref{dwallfig} also displays the magnetisation
$m$ for these domain wall solutions. It is interesting to observe that it is always positive, corresponding to a paramagnetic system, in contrast to the dyonic AdS-RN solutions
of pure Einstein-Maxwell theory which are diamagnetic  \cite{Hartnoll:2008kx,Denef:2009yy}
(see appendix A).

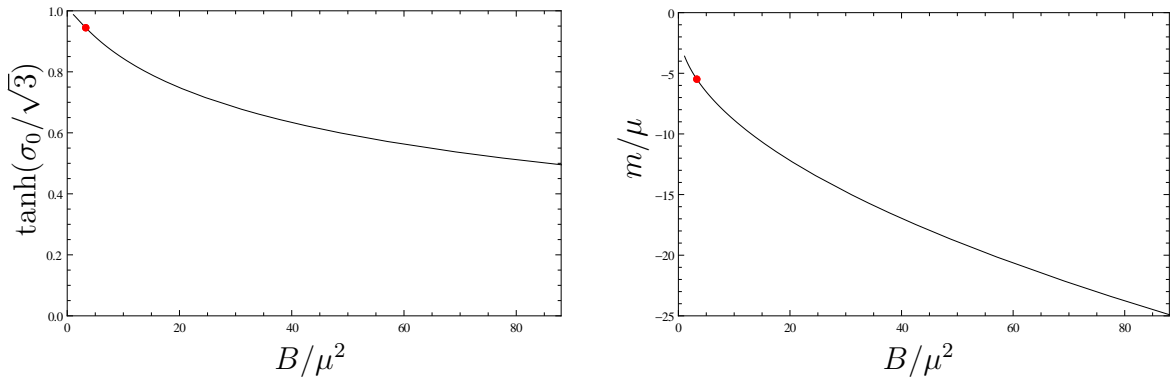
\begin{figure}[h!]
\begin{center}
\begin{picture}(0.1,0.25)(0,0)
\put(100,-10){\makebox(0,0){$B/\mu^2$}}
\put(-8,70){\makebox(0,0){\begin{sideways}{\rm tanh}($\sigma_0/\sqrt{3}$)\end{sideways}}}
\end{picture}\includegraphics[width=0.45 \textwidth]{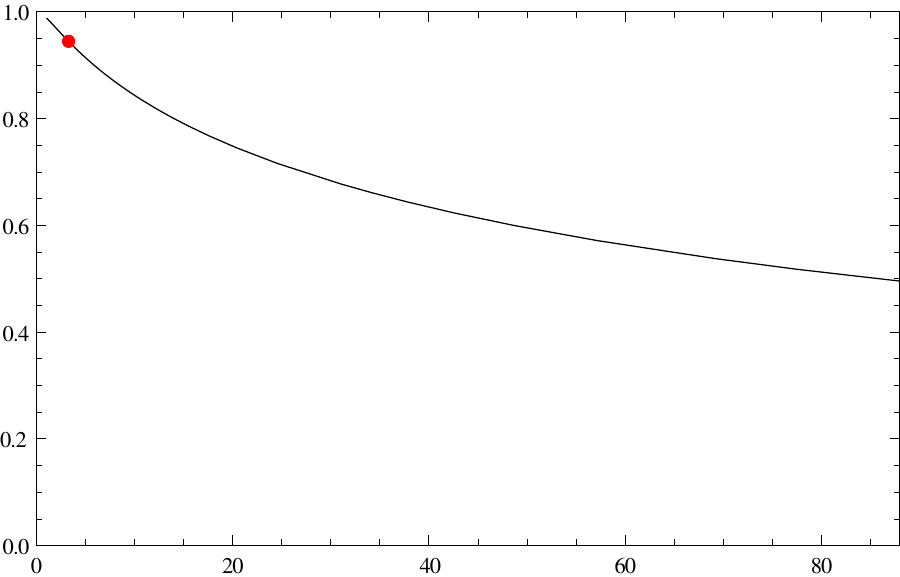}\qquad\quad
\begin{picture}(0.1,0.25)(0,0)
\put(100,-10){\makebox(0,0){$B/\mu^2$}}
\put(-8,70){\makebox(0,0){\begin{sideways}$m/\mu$\end{sideways}}}
\end{picture}\includegraphics[width=0.45
\textwidth]{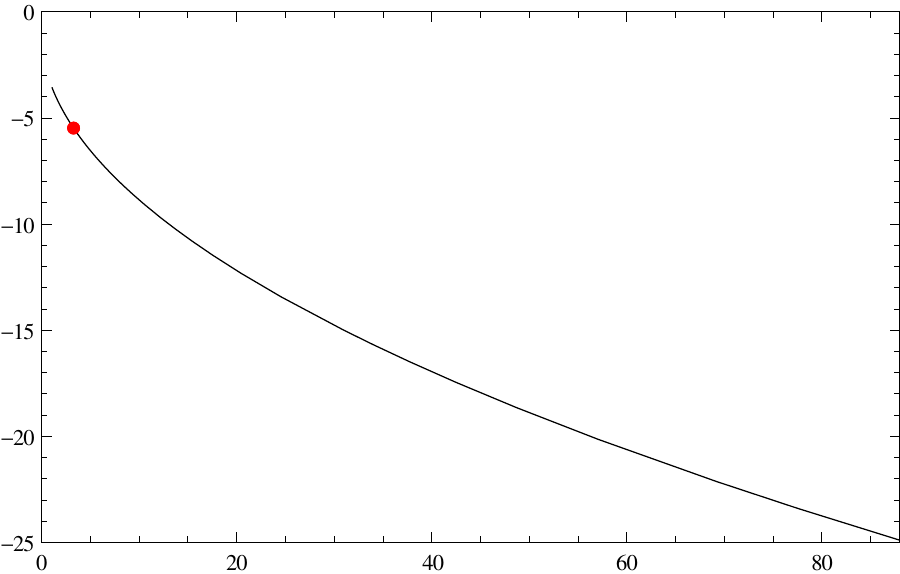}\vskip1em
\caption{Left panel: the one parameter family of domain wall solutions interpolating between $AdS_4$ in the UV, with deformation
data $(\mu,B)$, and dyonic $AdS_2\times\mathbb{R}^2$ solutions in the magnetic 
family in the IR, labelled by $\sigma_0$. For convenience
of presentation the vertical axis is given by $\tanh(\sigma_0/\sqrt{3})$. It is expected that
these domain walls exist for $B/\mu^2\to 0$ but that they have 
higher free energy than the domain walls in figure
\ref{dwallfig} for the same values of $B,\mu$. The red dots indicate superfluid instabilities, discussed in section \ref{secsfluid}, with the solutions being unstable to the left of the crosses. Right panel: a plot of the magnetisation $m/\mu$ as a function of $B/\mu^2$. Observe that the magnetisation is always negative corresponding to diamagnetism.
  \label{dwallfigmag}}
\end{center}
\end{figure}

We next consider domain wall solutions that approach $AdS_2\times\mathbb{R}^2$ solutions lying in the magnetic family. We display a one-parameter family in figure \ref{dwallfigmag} 
again with $B>0,\mu>0$. Note that the purely magnetic solution, the $T=0$ limit of the purely magnetically charged AdS-RN solution
\eqref{magadsrn}, is obtained when $\mu=0$
and hence $B/\mu^2\to\infty$ in the figure. 
Also note that we have only been able to stabilise the numerics for values of 
$B/\mu^2\gtrsim 1$ and in particular we have not obtained these
solutions for values of $B/\mu^2$ that overlap with the solutions in figure \ref{dwallfig}. It is most likely
that they do exist for all values of $B/\mu^2$ down to zero, but that they have higher free energy and hence are not thermodynamically 
relevant\footnote{This is based on two calculations.
Firstly, using the values for the free energy of the domain walls that we have constructed in figure \ref{dwallfigmag} and then extrapolating to smaller values of $B/\mu^2$. And secondly, we have also calculated some domain walls after adding in a deformation in the UV by the $\Delta=2$ operator dual
to $\sigma$ (see section \ref{fcomms}) where we were able to make a direct comparison of the free energies for the same UV data in some cases.}. We will continue with this assumption. 
Figure \ref{dwallfigmag} also displays the magnetisation
$m$ for these domain wall solutions. It is interesting to observe that it is always negative, corresponding to diamagnetism, in contrast to the domain wall solutions lying in the electric family of
figure \ref{dwallfig}.

\section{Dyonic Black hole solutions}\label{dybhs}

In this section we construct finite temperature black hole solutions that 
approach $AdS_4$ in the UV as in \eqref{expan}, again with $\sigma_1=\beta_a=0$. 
Also, as in the previous section, we choose $\mu>0$ and $B\ge 0$ with solutions with $B\le 0$
obtained using the symmetry \eqref{bsym}.

Our initial strategy is to heat up the $T=0$ domain wall solutions that we constructed in the previous section.
One focus of interest is what we will call ``region I'', with 
$0\le B/\mu^2\le (B/\mu^2)_I$. In this region, by definition, 
the unbroken phase is described by dyonic black holes whose zero temperature limit
is given by dyonic domain wall solutions approaching in the IR $AdS_2\times \mathbb{R}^2$ solutions lying
on the electric family. One might suspect that $(B/\mu^2)_I$ coincides with $(B/\mu^2)_{\rm max}$, the value of $B/\mu^2$ in which such domain walls
cease to exist (see figure \ref{dwallfig}). However, this is not quite true and in fact we have $(B/\mu^2)_I< (B/\mu^2)_{\rm max}$, 
because of the existence of a first order metamagnetic transition, as we shall explain.

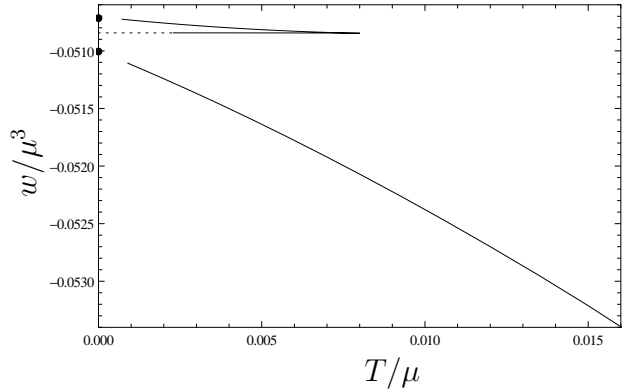
\begin{figure}[h!]
\begin{center}
\begin{picture}(0.1,0.25)(0,0)
\put(130,-10){\makebox(0,0){$T/\mu$}}
\put(-8,70){\makebox(0,0){\begin{sideways}$w/\mu^3$\end{sideways}}}
\end{picture}\includegraphics[width=0.5\textwidth]{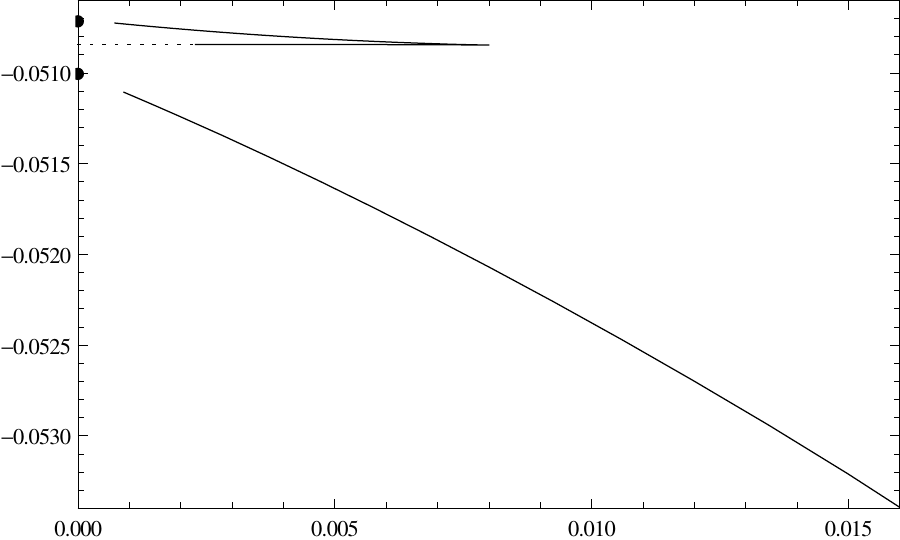}\vskip1em
\caption{A representative plot of the free energy of two families
of dyonic black hole solutions that exist in region I i.e. for values of $B/\mu^2$ with $0<B/\mu^2<(B/\mu^2)_I$.
At $T=0$ each family approaches a smooth domain wall solution interpolating between $AdS_4$ in the UV and
two different dyonic $AdS_2\times\mathbb{R}^2$ solutions in the IR. Notice that only the bottom branch
can be heated up to arbitrarily high temperatures; it is thermodynamically preferred and describes the unbroken phase.   
\label{RI}}
\end{center}
\end{figure}

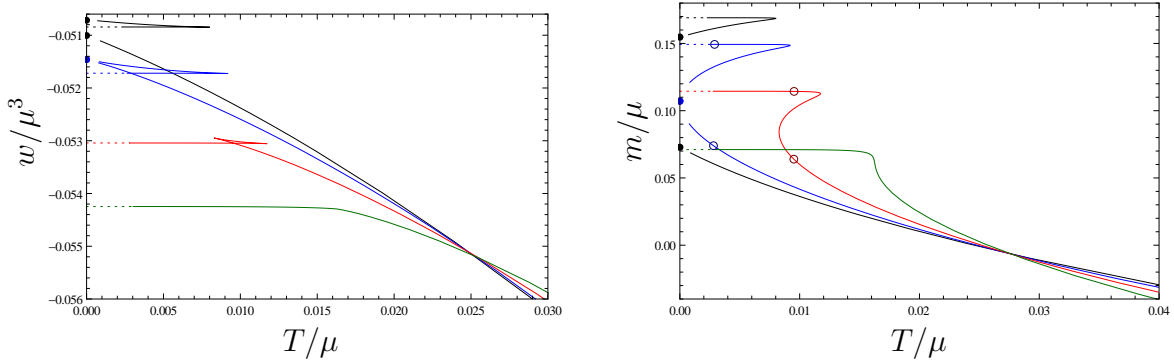
\begin{figure}[h!]
\begin{center}
\begin{picture}(0.1,0.25)(0,0)
\put(100,-10){\makebox(0,0){$T/\mu$}}
\put(-8,70){\makebox(0,0){\begin{sideways}$w/\mu^3$\end{sideways}}}
\end{picture}\includegraphics[width=0.45 \textwidth]{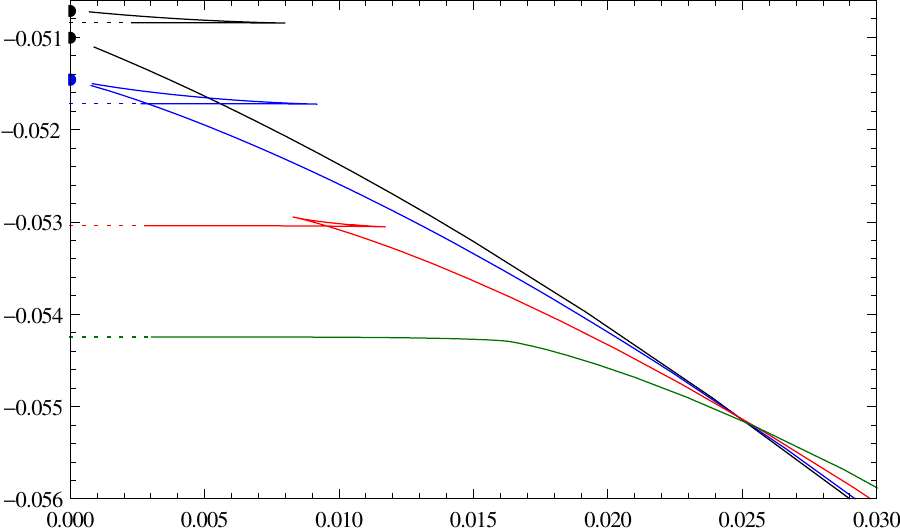}\qquad\quad
\begin{picture}(0.1,0.25)(0,0)
\put(100,-10){\makebox(0,0){$T/\mu$}}
\put(-8,70){\makebox(0,0){\begin{sideways}$m/\mu$\end{sideways}}}
\end{picture}\includegraphics[width=0.45\textwidth]{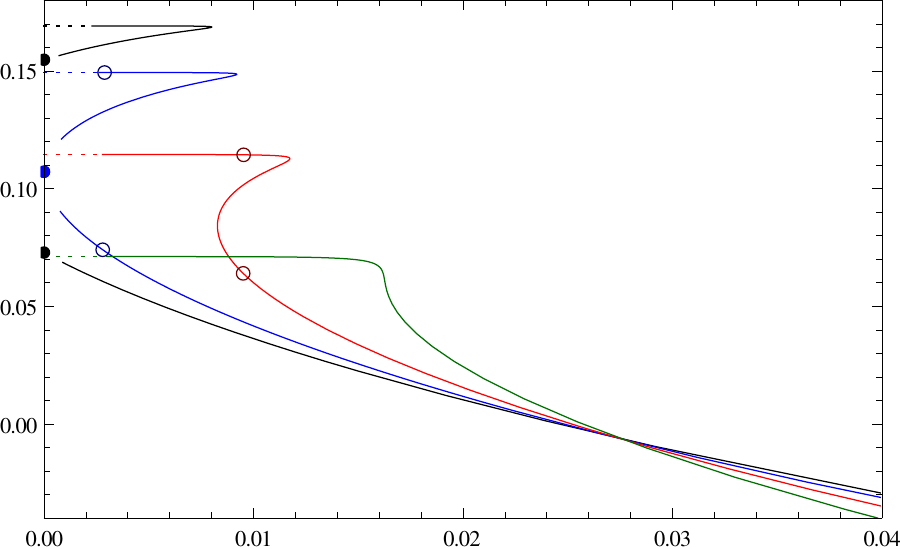}\vskip1em
\caption{A series of dyonic black holes for different values of $B/\mu^2$, with
increasing values of $B/\mu^2$ for the black, blue, red and green curves, respectively. The black curve
has $B/\mu^2<(B/\mu^2)_I$. The blue curve has $B/\mu^2= (B/\mu^2)_{\rm max}>(B/\mu^2)_I$ and the red and green curves have $B/\mu^2>(B/\mu^2)_{\rm max}$.
In the left and right plots we show the free energy and the magnetisation as a function of $T$. Observe that
for the blue and red curves there is a first order phase transition at finite temperature, marked with circles
on the right plot, where there is a discontinuous jump in the magnetisation. The solid black and blue dots refer to
domain wall solutions with dyonic $AdS_2\times\mathbb{R}^2$ solutions in the IR. 
The dashed 
curves indicate black hole solutions that are approaching hyperscaling violating behaviour in the 
IR as $T\to 0$.
  \label{seriesof}}
\end{center}
\end{figure}

In figure \ref{RI} we show a representative plot of the black hole solutions that we have constructed with $0<B/\mu^2<(B/\mu^2)_I$,
with $B/\mu^2$ close to $(B/\mu^2)_I$.
We can heat up the domain wall solution approaching the $AdS_2\times \mathbb{R}^2$ solution that lies on the upper branch of figure
\ref{dwallfig}
to arbitrary high temperatures. On the other hand for the domain wall solution approaching the $AdS_2\times \mathbb{R}^2$ solution that lies on the lower branch of figure \ref{dwallfig} we find that it can only be heated up to a maximum temperature before it goes back down to lower 
temperatures. Following this solution down to very low temperatures we find that the solution is becoming singular in the IR, approaching a hyperscaling violating behaviour as we discuss below.
We can see from figure \ref{RI} that the free energy for the branch of black holes that can be heated up to high temperatures is
always thermodynamically preferred and hence describes the high temperature unbroken phase of the system.
In the right plot of figure
\ref{seriesof} the black curves plot the magnetisation as a function of $T$ for the two black hole branches for the value
of $B/\mu^2$ given in figure \ref{RI} (the same value of $B/\mu^2$ for the black curve in the left plot of figure \ref{seriesof}). We observe that for
the thermodynamically preferred black holes the behaviour switches
from paramagnetism at low temperatures to diamagnetism at high temperatures.

Having discussed the black holes for a representative value of $B/\mu^2<(B/\mu^2)_I$ in region I, let us see what happens as we increase the value of $B/\mu^2$. A series of black hole solutions
is presented in figure \ref{seriesof}.
We see that at $B=B_{\rm max}$, corresponding to the blue curve, 
the two branches of black holes are coalescing at $T=0$ corresponding to the fact
that at $B/\mu^2=(B/\mu^2)_{\rm max}$ there is just a single domain wall solution mapping in the IR onto an $AdS_2\times \mathbb{R}^2$ solution
in the electric family. By examining the free energy, we see that the blue curve implies there is a first order phase
transition at finite temperature. We also see from the right panel in figure \ref{seriesof}, that there is an abrupt
change in the magnetisation and hence we have a first-order metamagnetic phase transition.
At low temperatures the preferred black holes, denoted by the dashed blue line in figure \ref{seriesof}, 
all exhibit hyperscaling violation in the IR, as we elaborate on further below.

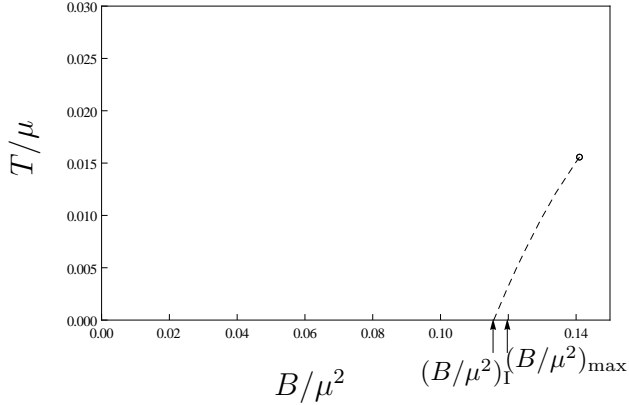
\begin{figure}[h!]
\begin{center}
\begin{picture}(0.1,0.25)(0,0)
\put(100,-10){\makebox(0,0){$B/\mu^2$}}
\put(159,-5){\makebox(0,0){{\footnotesize $(B/\mu^2)_{\rm I}$}}}
\put(197,0){\makebox(0,0){{\footnotesize $(B/\mu^2)_{\rm max}$}}}
\put(-8,80){\makebox(0,0){\begin{sideways}$T/\mu$\end{sideways}}}
\end{picture}\includegraphics[width=0.5 \textwidth]{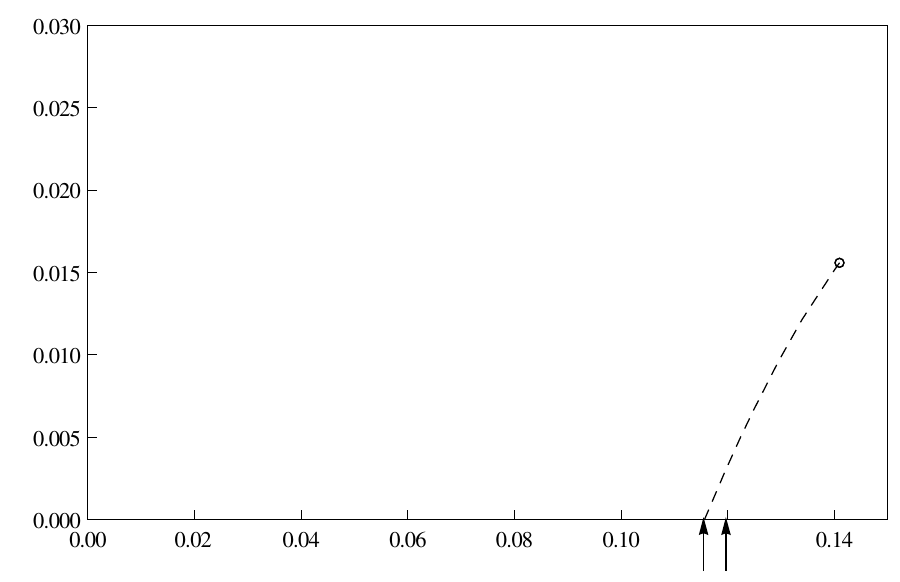}\qquad\quad\vskip1em
\caption{The phase diagram for the unbroken phase black holes. 
For Region I, with $0\le B/\mu^2<(B/\mu^2)_I$ 
the black holes at $T=0$ are domain walls that approach an $AdS_2\times\mathbb{R}^2$ solution in the electric family in the far IR. More precisely, they are domain walls that are on the upper part of the curve in the
left hand plot in figure \ref{dwallfig}.
For $(B/\mu^2)_I\le B/\mu^2\le (B/\mu^2)_{\rm max}$ 
similar black holes exist, but they are not thermodynamically preferred. Instead the dashed line represents
a line of first order phase transitions to a metamagnetic phase, ending in a second order critical point denoted by a round circle. For $B/\mu^2> (B/\mu^2)_I$, as $T\to 0$ the thermodynamically preferred
solutions exhibit hyperscaling violation in the IR with dynamical exponent $z=3/2$ and $\theta=-2$.
  \label{metmag}}
\end{center}
\end{figure}

For higher values of $B/\mu^2$, $B/\mu^2>(B/\mu^2)_{\rm max}$, illustrated by the red and the green curves in figure \ref{seriesof},
there is just a single branch of black holes all of which approach a singular solution at $T=0$. For the red curve
we again see that there is a first order phase transition at finite temperature with the transition temperature depending on $B/\mu^2$. Increasing $B/\mu^2$ further we get to the green curve and the first order transition comes to an end at a critical point which is second (or higher) order.
In the right plot of figure \ref{seriesof} we see that the jump in the magnetisation, present for the first order transitions,
is decreasing as one increases $B/\mu^2$, disappearing for the green curve. 
In figure \ref{metmag} we have summarised the $T,B$ phase diagram for these unbroken phase black holes.

\subsection{Hyperscaling violation}
\begin{figure}[h!]
\begin{center}
\begin{picture}(0.1,0.25)(0,0)
\put(159,-5){\makebox(0,0){{\footnotesize $\log\, T/\mu$}}}
\put(-8,100){\makebox(0,0){\begin{sideways}$\log\, s/\mu^2$\end{sideways}}}
\end{picture}
\includegraphics[width=0.7 \textwidth]{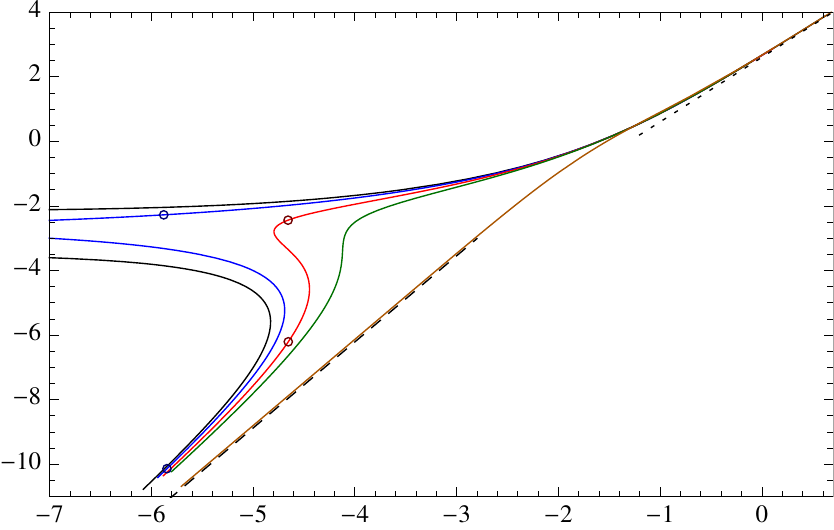}
\vskip1em
\caption{A plot of $\log\, s/\mu^2$ versus $\log\, T/\mu$ for the dyonic black hole solutions with different values of $B/\mu^2$.
The black, blue, red and green curves are the same black holes as in figure \ref{seriesof} and the brown curve
is a black hole with a much larger value of $B/\mu^2$. As $T\to \infty$ the dotted line corresponds
to $s\propto T^2$, associated with the $AdS_4$ asymptotics in the UV. 
As $T\to 0$ the dashed line corresponds
to $s\propto T^{8/3}$, associated with the hyperscaling violating asymptotics in the IR with $z=3/2$ and
$\theta=-2$.
  \label{both scale}}
\end{center}
\end{figure}

We now return to the properties of the solutions denoted by dashed curves in figure \ref{seriesof}
in the limit that $T\to 0$. We first observe from the free-energy curves in
figure \ref{seriesof} that as $T\to 0$ they all have $\partial_T w\to 0$ and hence the entropy density $s\to 0$.
In figure \ref{both scale} we have plotted the behaviour of $\log s$ versus $\log T$ which clearly reveals, especially for large values
of $B/\mu^2$, that $s\propto T^{8/3}$ as $T\to 0$. This strongly suggests that the solutions all exhibit an emergent
scaling behaviour in the IR as $T\to 0$. 

The scaling behaviour can be identified by returning to the equations of motion coming from \eqref{actansatz}.
In particular, we have found the following one-parameter family of solutions, at leading order in $r$ as $r\to 0$,
given by
\begin{align}\label{hsvsol}
ds^2&=-r^{5/2} dt^2 +\frac{dr^2}{2L r}+r^2(dx_1^2+dx_2^2)\,,\nn
F&=Bdx_1\wedge dx_2\,,\nn
\sigma&=\sigma_0+\sqrt{3}\log r\,,
\end{align}
where $\sigma_0$ is a constant and
\begin{align}
B=\pm \frac{2}{\sqrt 3}e^{-\frac{2\sigma_0}{\sqrt 3}},
\quad L=\frac{32}{33}e^{-\frac{\sigma_0}{\sqrt 3}}\,.
\end{align}
This purely magnetic configuration solves the equations of motion to leading order as $r\to 0$, and moreover, 
we find that the $F\wedge F$ coupling in \eqref{action} is not playing a role.
A simple co-ordinate transformation $r\to (L/2)\rho^{-2}$, combined with a rescaling of the
time and spatial co-ordinates, reveals this to be a hyperscaling violating metric 
\cite{Charmousis:2010zz,Ogawa:2011bz,Huijse:2011ef} given by
\begin{align}
ds^2=\rho^{-(2-\theta)}(-\rho^{-2(z-1)}d t^2+d\rho^2+d\bar x_i d x_i)\,,
\end{align}
with dynamical exponent $z=3/2$ and hyperscaling violation exponent $\theta=-2$. In particular, under the scaling
\begin{align}
t\to \lambda^z t,\quad x_i \to \lambda x_i,\quad \rho \to \lambda \rho\,,
\end{align}
the metric scales as $ds\to \lambda^{\theta/2} ds$. If one heats up this class of
hyperscaling violating solutions one finds that the entropy density behaves as $s\propto T^{(2-\theta)/z}=T^{8/3}$,
which is exactly the same behaviour we see for our solutions as $T\to 0$.

A more detailed look at our numerical solutions provides
additional evidence that as $T\to 0$ the solutions are domain walls interpolating between the hyperscaling
violating behaviour \eqref{hsvsol} in the IR and $AdS_4$ in the UV. We leave the detailed construction of such domain wall solutions to future work, but we note one final point.
As $T\to 0$ the black hole solutions have  $\sigma_+\to -\infty$ at the event horizon 
and also the UV charge $q$ is given by $q\to B$. A consideration of \eqref{chgecon} shows that 
the origin of the electric charge in the far IR for these $s=0$ ground states is arising purely from the $F\wedge F$
coupling in \eqref{action}. 

\subsection{Other dyonic black hole solutions}
The alert reader will have noticed that in addition to the two branches of black holes presented in figure \ref{RI}, mapping onto
a $T=0$ domain wall solution approaching an $AdS_2\times \mathbb{R}^2$ solution in the electric family, there could be an
additional branch of black holes which maps onto a $T=0$ domain wall solution approaching an $AdS_2\times \mathbb{R}^2$ solution
in the magnetic family as in figure \ref{dwallfigmag}. Indeed we have constructed such solutions for values of 
$B/\mu^2$ as in figure \ref{dwallfigmag}, which we recall are much higher than those in figure \ref{RI}. By extrapolating 
the free energy of these solutions down to smaller values of $B/\mu^2$, combined with the observations that we made
in footnote 6, we expect these black holes are never thermodynamically preferred over those 
presented in figure \ref{RI} and hence they will not change the picture summarised in figure \ref{metmag}.
We also note that the solutions of this type that we have constructed exist up to a maximum
temperature and then return to low temperatures, approaching a singular solution as $T\to 0$, analogous to
the upper branch in figure \ref{RI}. In fact these solutions also appear to approach the hyperscaling
violating solutions given in \eqref{hsvsol}.

\section{Superfluid and striped instabilities}
\label{sasi}
In section 7.1 we analyse striped instabilities of the dyonic $AdS_2\times\mathbb{R}^2$ solutions in both
the electric and magnetic families
given in section \ref{axdysol}. This allows us to deduce the existence of instabilities of the corresponding domain walls constructed in section \ref{ddws} and the thermodynamically preferred unbroken phase black hole solutions in region I,
i.e. with 
$0\le B/\mu^2<(B/\mu^2)_I$, that map onto domain walls at zero temperature, constructed in section \ref{dybhs} (see figure \ref{metmag}). 
Note that the critical temperatures for the
existence of the striped instabilities for the black holes
is very low and we have not been able to stabilise the numerics to find their precise values.
In section 7.2  we construct the superfluid instabilities both for the dyonic $AdS_2\times\mathbb{R}^2$ solutions
and for the thermodynamically preferred unbroken phase black holes in region I and we find that they only
exist for $0\le B/\mu^2\le (B/\mu^2)_c<(B/\mu^2)_I$. 
In section 7.3 we discuss the implications for the full phase diagram of the system, which we summarised in figure
\ref{TBphase}.

\subsection{Striped instabilities}
We consider spatially modulated, or ``striped", perturbations about the dyonic $AdS_2\times\mathbb{R}^2$
solutions given in section \ref{axdysol}. We write these solutions as 
\begin{align}\label{ads2sols2}
ds^2&=L^2\left(-\rho^2 dt^2+\rho^{-2}d\rho^2\right)+ dx_1^2+dx_2^2\,,\nn
A&=-EL^2\rho dt -\tfrac{1}{2}Bx_2 dx_1+\tfrac{1}{2}Bx_1 dx_2,\qquad\sigma=\sigma_0
\end{align}
with $\sigma_0$, $E,L$ and $B$ constants satisfying \eqref{con1},\eqref{con2}. We consider the linearised
perturbation
\begin{align}
\delta g_{tt}(\rho,x_1)&=L^2\rho^2 h_{tt}(\rho)\cos(k x_1)\,,\nn
\delta g_{tx_2}(\rho,x_1)&=L^2\rho h_{tx_2}(\rho)\sin(k x_1)\,,\nn
\delta g_{x_1x_1}(\rho,x_1)&= h_{x_1x_1}(\rho)\cos(k x_1)\,,\nn
\delta g_{x_2x_2}(\rho,x_1)&=h_{x_2x_2}(\rho)\cos(k x_1)\,,\nn
\delta A_t(\rho,x_1)&=\rho\delta a_t(\rho)\cos(kx_1)\,, \nn
\delta A_{x_2}(\rho,x_1)&=\delta a_{x_2}(\rho)\sin(kx_1)\,, \nn
\delta \sigma(\rho,x_1)&=s(\rho)\cos(kx_1)\,,
\end{align}
where the wave-number $k$ is a constant.
Substituting into the equations of motion we are lead to a system of coupled linear differential equations that are
second order in
$h_{tt}(\rho)$, $h_{tx_2}(\rho)$, $a_t(\rho)$, $a_{x_2}(\rho)$, $s(\rho)$
and, when $k\ne 0$, first order in $h_{x_1x_1}(\rho)$, $h_{x_2x_2}(\rho)$. When $k=0$ they are first order in $h_{x_2x_2}(\rho)$
but second order in $h_{x_1x_1}(\rho)$.

We are interested in analysing the spectrum of scaling dimensions of the operators in the CFT dual to the $AdS_2$ solution
and in particular whether there are any modes that violate the BF bound. We therefore look for solutions where the seven functions
of $\rho$ are of the form
${\bf v} \rho^{-\delta}$ where ${\bf v}$ is a constant vector and $\delta$ is a constant related to the scaling dimension via $\Delta =\delta$ or $\Delta=1-\delta$. The BF bound is given by $\Delta=1/2$. The system of equations then takes the form
 ${\bf M v}=0$ where ${\bf M}$ is a $7\times 7$ matrix. Demanding that non-trivial ${\bf v}$ exists implies that det${\bf M}=0$
 and this specifies the possible values of $\delta$ as a function $k$. 

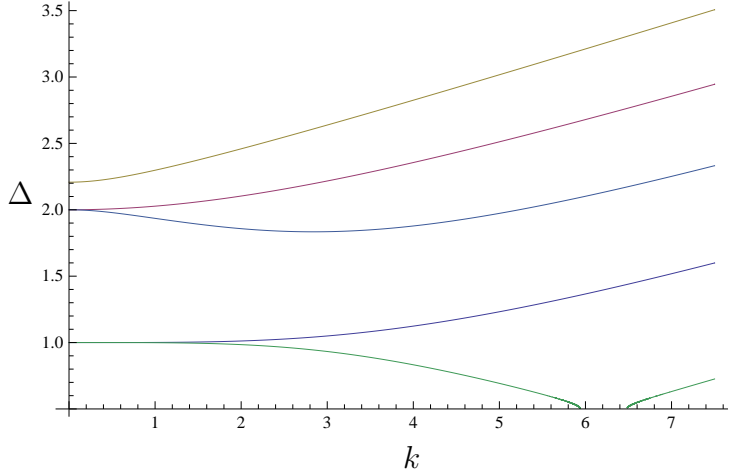
\begin{figure}[h!]
\begin{center}
\begin{picture}(0.1,0.25)(0,0)
\put(140,-10){\makebox(0,0){$k$}}
\put(-8,90){\makebox(0,0){$\Delta$}}
\end{picture}\includegraphics[width=0.6\textwidth]{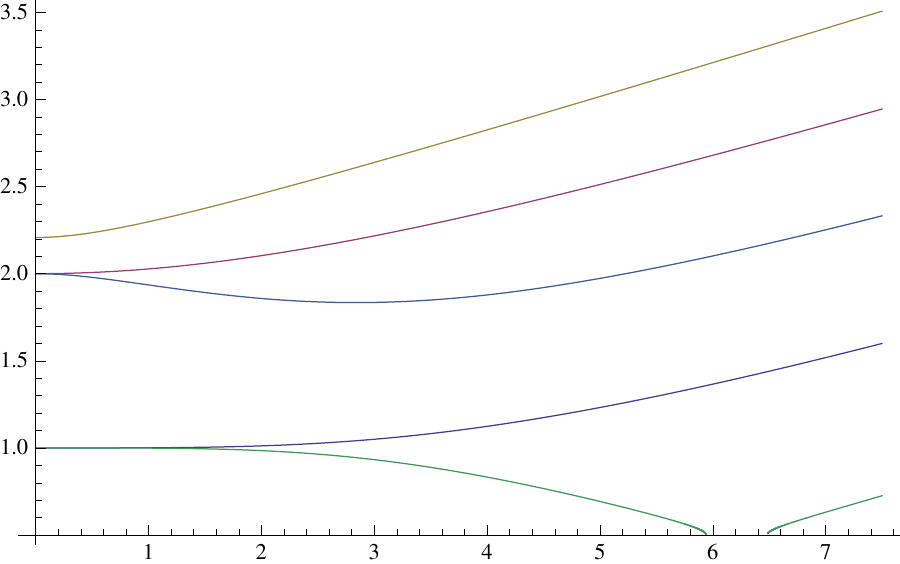}\vskip1em
\caption{The spectrum of scaling dimensions $\Delta$ for spatially modulated striped modes of the purely electric $AdS_2\times\mathbb{R}^2$ 
solution as a function of wave-number $k$. For $k\in(5.94,6.48)$ the BF bound is violated and there are
striped instabilities. After rescaling the momentum we find exactly the same spectrum for the entire family of dyonic $AdS_2\times\mathbb{R}^2$ 
solutions.
  \label{specelfam}}
\end{center}
\end{figure}

We first focus on the electric family of $AdS_2\times\mathbb{R}^2$ solutions. We further restrict to the purely electric solution with
$\sigma_0=B=0$. 
We find that the spectrum of $\delta$ is symmetric with respect to $k\to -k$ and we have displayed the values of
$\Delta$ in figure \ref{specelfam} for $k\ge 0$. 
In particular, we find that, for $k\ge 0$,
the BF bound is violated in the approximate range $k\in(5.94,6.48)$, as already observed in \cite{Donos:2011bh}.
We next analyse what happens as we move along the electric family of dyonic solutions by allowing $\sigma_0\ne 0$. We find the surprising result that
after rescaling $k\to \sqrt{\cosh(\sigma_0/\sqrt{3})}k$ the spectrum is independent of $\sigma_0$. It would be interesting to better understand the
underlying origin of this result. Moving to the magnetic family of $AdS_2\times\mathbb{R}^2$ solutions, we find that the spectrum is identical to the electric family, which is presumably a consequence of the duality symmetry \eqref{dualtrans}. 

We conclude that all the dyonic $AdS_2\times\mathbb{R}^2$ solutions of section \ref{axdysol} have spatially modulated instabilities.

\subsection{Superfluid instabilities}\label{secsfluid}
To study superfluid instabilities we now consider the consistent Kaluza-Klein truncation of \cite{Gauntlett:2009bh} which also includes
the charged scalar field, denoted by $\chi$ in \cite{Gauntlett:2009bh}. Writing $\chi=\sqrt{\tfrac{4}{3}}\tanh(\eta/2)e^{2i\theta}$, where $\eta,\theta$ are real, 
the action of \cite{Gauntlett:2009bh} can be written
\begin{align}\label{action2}
S= &\frac{1}{16\pi G} \int d^4 x \sqrt{-g} \Bigg(R - \frac{1}{2}(\partial \sigma)^2  - \frac{\tau(\sigma)}{4} F^2 
- V(\sigma,\eta)\nn
&\qquad\qquad-\frac{1}{2} (\partial\eta)^2-  2\sinh^2\eta(\partial\theta-A)^2
\Bigg)
+
\frac{1}{32\pi G}\int \vartheta(\sigma) F\wedge F\,,
\end{align}
where $\tau(\sigma)  =\frac{1}{{\cosh\sqrt{3}\sigma}}$ and
$\vartheta(\sigma)  = \tanh\sqrt{3}\sigma$ as before, and now
\begin{align}
V(\sigma,\eta) & \equiv   -24\cosh\left({\frac{\sigma}{\sqrt{3}} }\right)\cosh^4\left(\frac{\eta}{2}\right)
\left[1  -\frac{4}{3}\tanh^2\left(\frac{\eta}{2}\right)\cosh^2\left({\frac{\sigma}{\sqrt{3}}}\right)\right]  \,.
\end{align}

We want to consider linearised fluctuations of the charged scalar field about the dyonic solutions with $\eta=0$,
that we have constructed earlier. Similar computations were first carried out for other models in 
\cite{Albash:2008eh,Hartnoll:2008kx}.
We can consistently work in a gauge with $\theta=0$. The linearised equation of motion
for $\eta$ then reads
\begin{align}
[\nabla^2 -4A^2-f(\sigma)]\eta=0\,,
\end{align}
where $\nabla$ and $A$ refer to the background solution and
\begin{align}
f(\sigma)=4\cosh(\sqrt{3}\sigma)-12\cosh\frac{\sigma}{\sqrt{3}}\,.
\end{align}

First consider perturbations about the dyonic $AdS_2\times\mathbb{R}^2$ solutions as given in 
\eqref{ads2sols2}. We focus on the lowest Landau level by
writing
\begin{align}
\eta(t,\rho,x_i)=e^{-\frac{|B|}{2}(x_1^2+x_2^2)}\bar\eta(t,\rho)\,.
\end{align}
We then find that
$\bar\eta$ satisfies
\begin{align}
[\nabla^2_{AdS_2}-L^2M^2]\bar\eta=0, \quad\quad M^2=2|B|+f(\sigma_0)-4E^2 L^2\,,
\end{align}
where $\nabla^2_{AdS_2}$ is the Laplacian for a unit radius $AdS_2$. This mode violates the $AdS_2$ BF bound when
$L^2M^2<-1/4$. Focussing on $\sigma_0\ge 0$, for the electric family we find that the mode is unstable apart
from the range of solutions where, approximately, $|\sigma_0|\in(1.01, 2.94)$.
We also find that for the magnetic family of solutions this mode violates the BF bound for (approximately) $|\sigma_0|\ge3.08$.
In section \ref{ddws} we constructed domain walls for UV data $\mu,B$ which map onto domain walls
that approach dyonic $AdS_2\times\mathbb{R}^2$ solutions on either the electric or magnetic family in the IR.
In figures \ref{dwallfig} and \ref{dwallfigmag}
we have indicated which of these domain wall solutions must be unstable.

We can also consider linearised perturbations about the dyonic black hole solutions lying in the ansatz \eqref{solans} that we constructed in
section \ref{dybhs}. In particular, we want to consider static, normalisable modes which appear at a critical temperature at which the superfluid instability appears. At this critical temperature new superfluid black hole solutions will appear. The mode that will have the highest critical temperature has the form
\begin{align}
\eta=e^{-\frac{|B|}{2}(x_1^2+x_2^2)}R(r)\,,
\end{align}
where $R(r)$ satisfies the ODE
\begin{align}
r^{-2}e^{\beta/2}\left(r^2 e^{-\beta/2}g R'\right)'-\left(\frac{2|B|}{r^2}+f(\sigma)-4\phi^2e^{\beta}g^{-1}\right)R=0\,.
\end{align}
At the black hole event horizon we have that $R(r_h)$ is a constant and we can use the linearity of the ODE to set this to unity. 
At the $AdS_4$ boundary we demand that 
$R(r)=\eta_1/r+\eta_2/r^2+\dots$ with $\eta_1=0$, corresponding to a spontaneous breaking of the abelian symmetry.
We thus have fixed two integration constants for the second order ODE and hence we expect that solutions will appear at specific temperatures. 

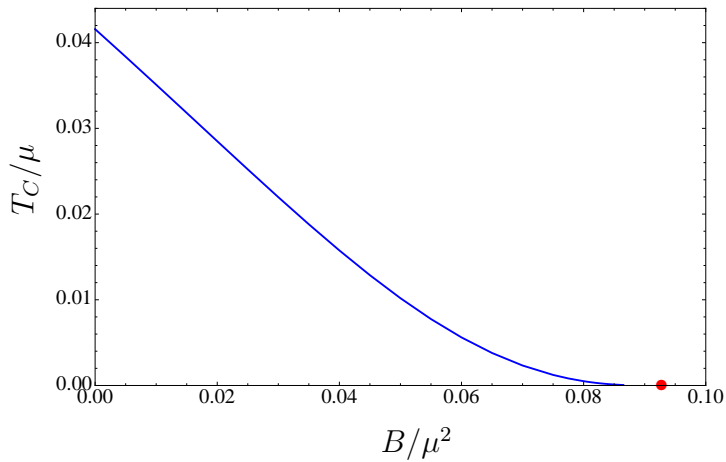
\begin{figure}[h!]
\begin{center}
\begin{picture}(0.1,0.25)(0,0)
\put(140,-10){\makebox(0,0){$B/\mu^2$}}
\put(-8,90){\makebox(0,0){\begin{sideways}$T_C/\mu$\end{sideways}}}
\put(232.6,12){{\color{red}\circle*{4}}}
\end{picture}\includegraphics[width=0.6\textwidth]{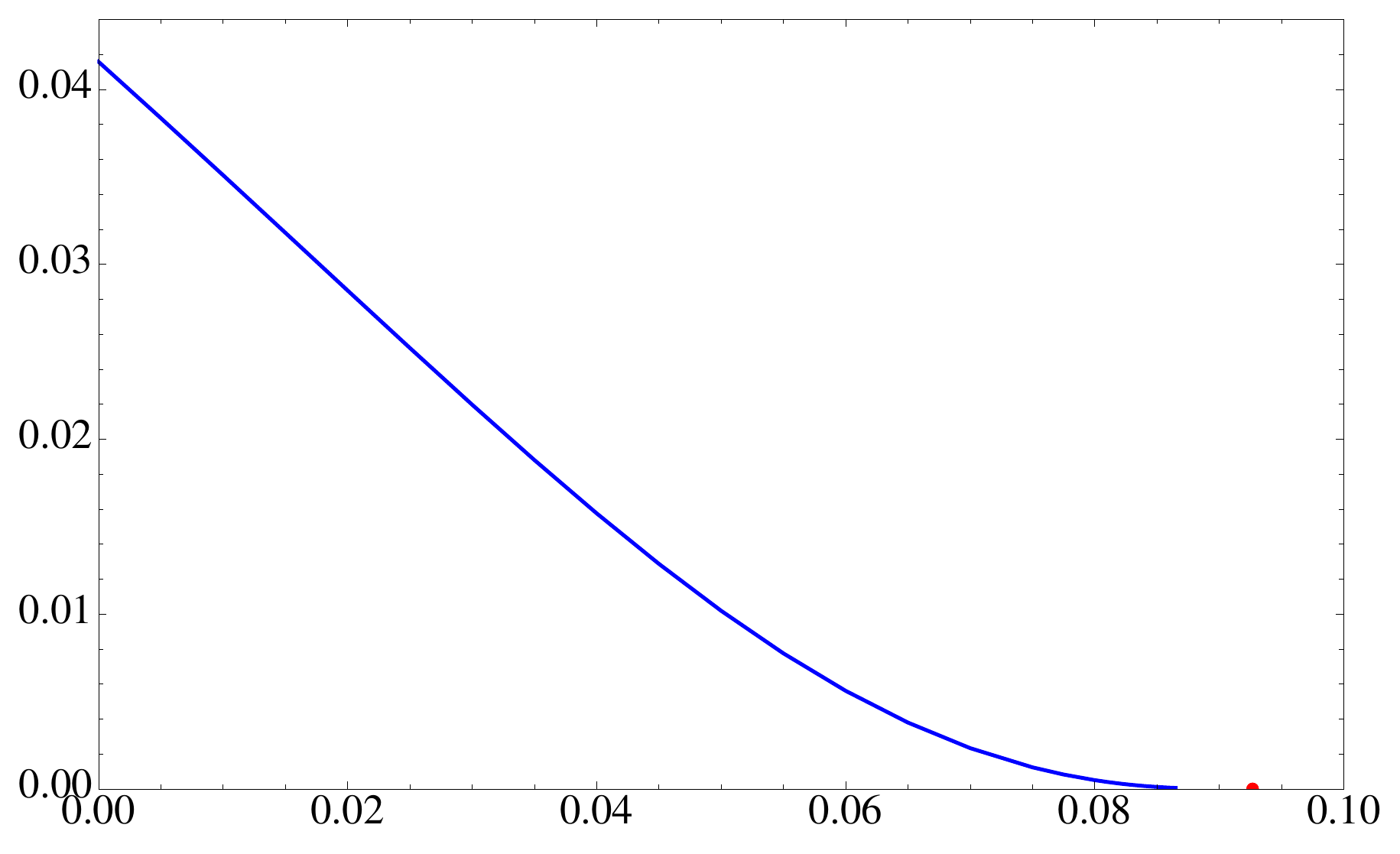}\vskip1em
\caption{A plot of the critical temperature at which the superfluid phase transition sets in, for the thermodynamically
preferred normal phase black holes summarised in figure \ref{metmag}
with $B/\mu^2<(B/\mu^2)_I$. The red dot corresponds to the red dot
appearing on the upper branch in the left hand plot of figure \ref{dwallfig}.
\label{sftb}}
\end{center}
\end{figure}

We are most interested in superfluid instabilities that appear on the thermodynamically preferred dyonic black hole solutions that we summarised in figure \ref{metmag}. The temperature at which the normalisable mode appears as a function of $B/\mu^2$ are presented  in figure \ref{sftb} and we see that they exist in the range $0\le B/\mu^2\le (B/\mu^2)_c$ with 
$(B/\mu^2)_c<(B/\mu^2)_I$. We expect that $(B/\mu^2)_c$ is the same limiting value at which the thermodynamically preferred
domain wall solutions on the electric branch have an instability in the $AdS_2\times\mathbb{R}^2$ region in the IR; the latter is marked with a red dot in figure \ref{sftb}.

\subsection{Conclusions about the full phase diagram}
In region I, with $0\le B/\mu^2<(B/\mu^2)_I$, the thermodynamically preferred unbroken phase black holes approach
a domain wall solution at zero temperature that maps onto a dyonic $AdS_2\times\mathbb{R}^2$ solution in the electric family
at $T=0$, as in figure \ref{metmag}. More precisely, they are domain walls that are on the upper part of the curve in 
the left hand plot in figure \ref{dwallfig}.
For $0\le B/\mu^2\le (B/\mu^2)_c<(B/\mu^2)_I$ there are both superfluid and striped instabilities
with the critical temperature for the superconducting instability monotonically decreasing to zero at $B/\mu^2=(B/\mu^2)_c$.
For $(B/\mu^2)_c\le B/\mu^2\le (B/\mu^2)_I$ there are no longer superfluid instabilities but the striped instabilities persist.
The critical temperature for the striped instabilities is very low for all values of $B/\mu^2$.

In order to deduce the phase diagram one needs to construct the fully back reacted black hole solutions for
both the superfluid and the striped black holes. While for $B=0$ the back-reacted superfluid black holes
were constructed by solving ODEs \cite{Gauntlett:2009dn}, for $B\ne 0$ we need to solve PDEs\footnote{It is likely that 
they form some kind of a vortex lattice e.g. \cite{Maeda:2009vf}. One can speculate that at least for small $B$ at $T=0$ they become domain walls interpolating
between two $AdS_4$ spaces, in order to match what happens at $B=0$ \cite{Gauntlett:2009dn}.}.
Similarly, constructing the back reacted striped black holes will also require solving PDEs, and this will be especially
challenging because of the low critical temperature at which they appear.

To proceed, we therefore make some reasonable simplifying assumptions. 
Firstly, that all of the superfluid and striped black holes
arise as second order transitions from the branch of unbroken phase black holes and secondly that they then continue
down to $T=0$ without sprouting additional branches. 
Finally,
we also assume that the free energy curves only
cross at most once as one lowers the temperature. To deduce the phase diagram of figure \ref{TBphase} then just requires a little
thought concerning the free energy of the black holes.
 
\begin{figure}[h!]
\begin{center}
\begin{picture}(0.1,0.25)(0,0)
\put(120,-5){\makebox(0,0){$T/\mu^2$}}
\put(-8,90){\makebox(0,0){\begin{sideways}$w/\mu^3$\end{sideways}}}
\end{picture}\includegraphics[width=0.45\textwidth]{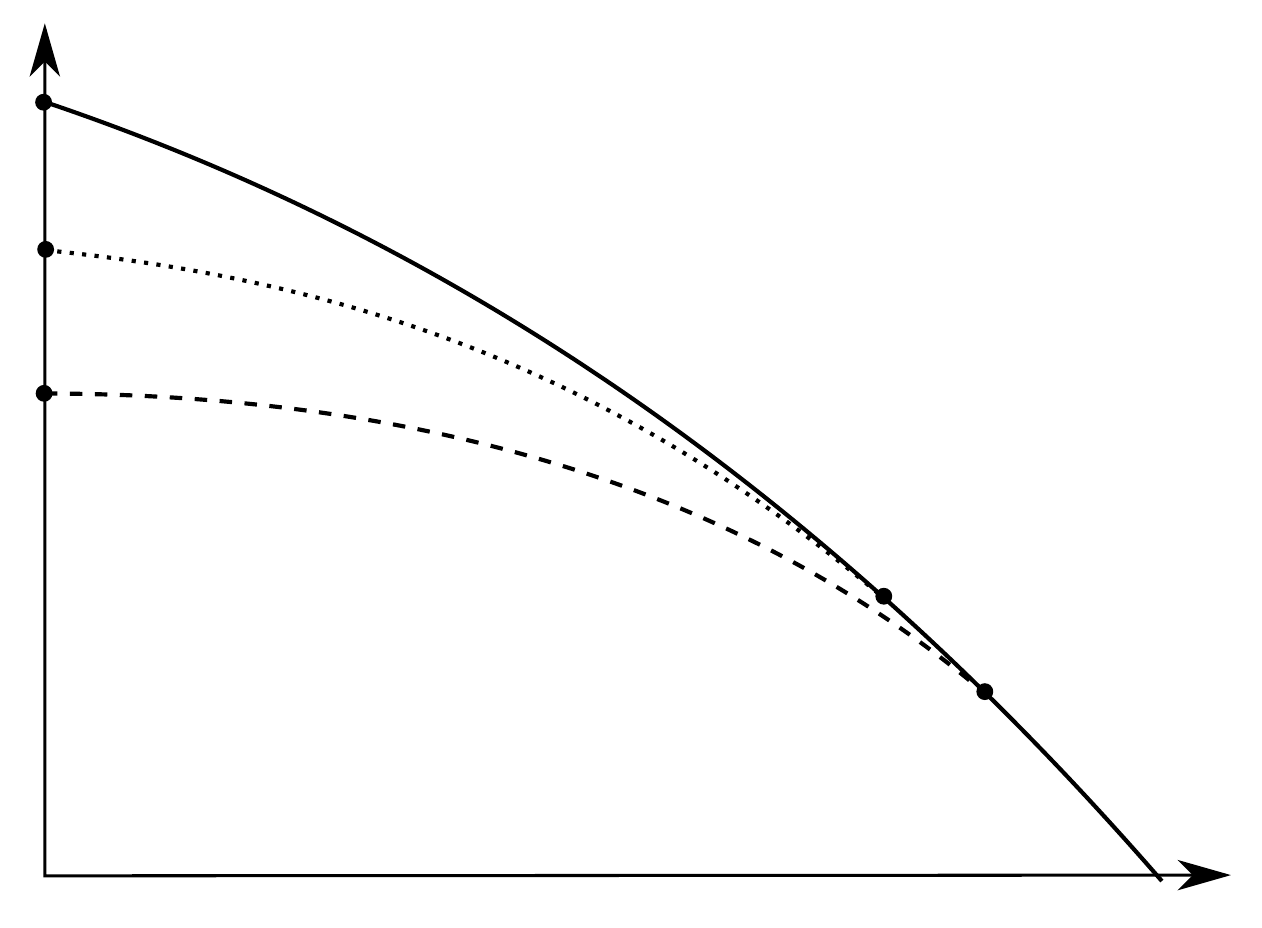}\quad
\begin{picture}(0.1,0.25)(0,0)
\put(120,-5){\makebox(0,0){$T/\mu^2$}}
\put(-8,90){\makebox(0,0){\begin{sideways}$w/\mu^3$\end{sideways}}}
\end{picture}\includegraphics[width=0.45\textwidth]{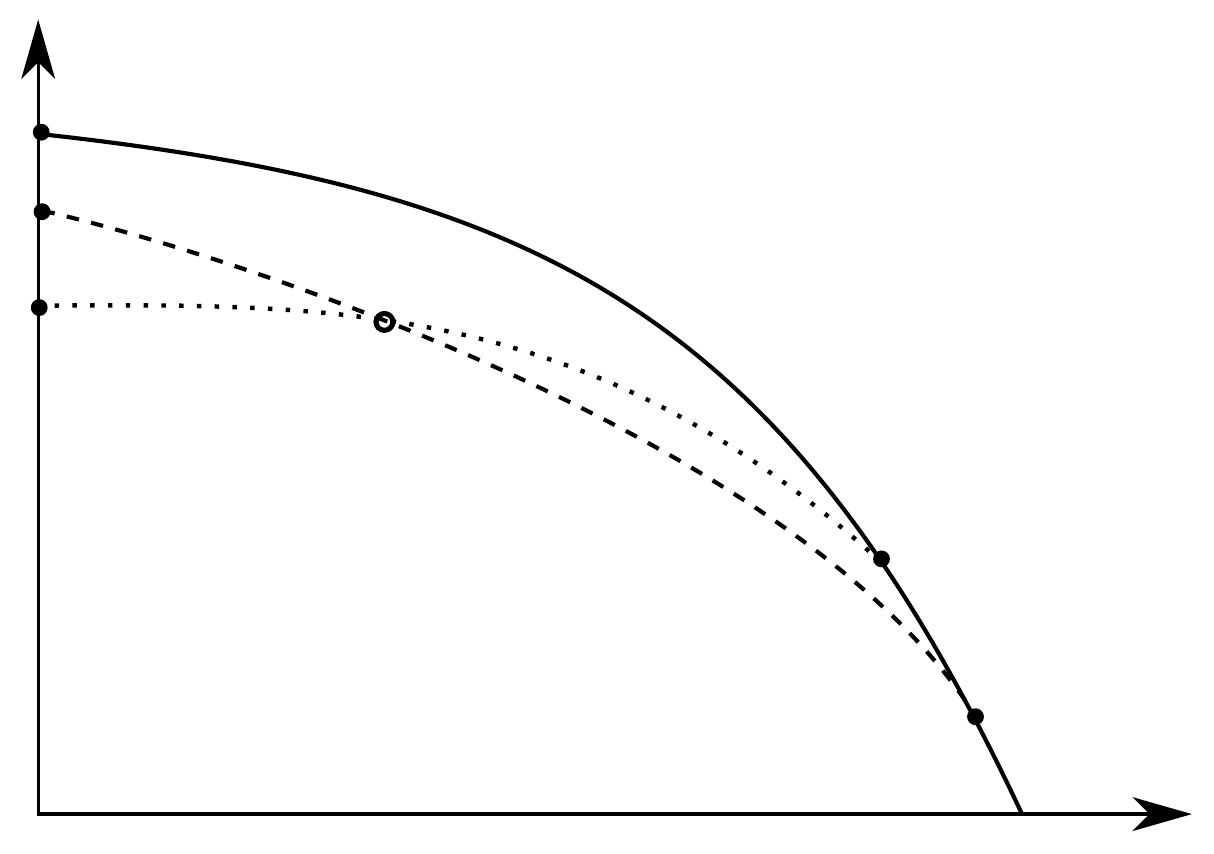}\quad\vskip1em
\caption{Schematic behaviour of the free energy versus temperature plots for
the unbroken phase black holes (solid lines), the superfluid black holes (dashed lines)
and the striped black holes (dotted lines). Figure (a) corresponds to values of $B/\mu^2$ with
$0\le B/\mu^2\le (B/\mu^2)_{(i)}$ in figure \ref{TBphase}. Figure (b) corresponds to $B/\mu^2$ being slightly bigger than $(B/\mu^2)_{(i)}$
in figure \ref{TBphase}. One could also have the slightly alternative scenario where the roles of the superfluid and striped branches are interchanged in the crossover. \label{schemefe}}
\end{center}
\end{figure}
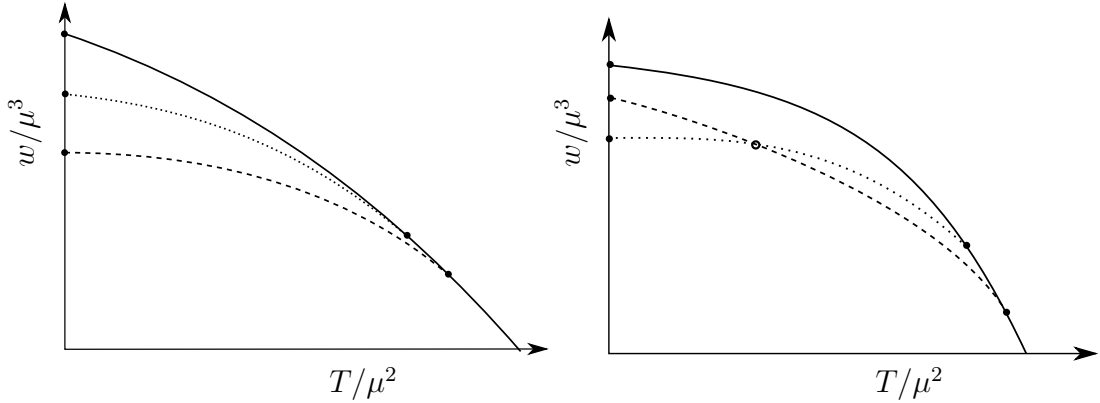

For small $B/\mu^2$ 
we stay on the superfluid branch all the way down to $T=0$ as shown in figure \ref{schemefe} (a) 
and also in figure \ref{TBphase}. For $(B/\mu^2)_c<B/\mu^2<(B/\mu^2)_I$ there
is no longer a branch of superfluid black holes and hence we must have a striped phase down to zero temperature,  
as depicted in figure \ref{TBphase}. What happens near $B/\mu^2=(B/\mu^2)_c$? One possibility is depicted in figure \ref{schemefe} (b): one first moves
onto the superfluid phase via a second order transition and then onto the striped phase via a first order transition. This possibility
is adopted in the phase diagram of figure \ref{TBphase} along with the tri-critical point that must necessarily appear and the first order transition
at $T=0$ at $B/\mu^2=(B/\mu^2)_{(i)}$.
A slightly different possibility is that  in figure \ref{TBphase} $(B/\mu^2)_{(i)}$ is closer to $(B/\mu^2)_c$ so that one would move from a striped phase via a first-order transition into a superfluid phase.
At the interface of stripes and metamagnetism a very similar kind of reasoning, which we won't spell out implies that the striped black holes will exist for small temperatures up to
$(B/\mu^2)_{(ii)}$ with $(B/\mu^2)_{(ii)}>(B/\mu^2)_I$ and generically $(B/\mu^2)_{(ii)}<(B/\mu^2)_{\rm max}$.

\section{Final Comments}
\label{fcomms}
We have shown that the phase structure of the $d=3$ CFTs, as a function of $T,\mu, B$, exhibit a rich phenomenology, summarised in figure \ref{TBphase}. It would be very interesting to identify
the $T=0$ ground states that emerge in the superfluid and striped phases with $B\ne 0$. However, this appears to be
a very challenging problem at the technical level. On the other hand we have successfully identified the $T\to 0$
hyperscaling violating ground states, with $z=3/2$, $\theta=-2$, 
for larger values of the magnetic field. It would be desirable to 
directly construct the $T=0$ domain wall solutions interpolating between the hyperscaling violating solutions
\eqref{hsvsol} in the IR and $AdS_4$ in the UV.

All of the black hole solutions in this paper are not deformed in the UV by the operator ${\cal O}_\sigma$ of dimension $\Delta=2$, dual to the pseudo-scalar $\sigma$. Recall that the role of such deformations on the superfluid instability when $B=0$ were analysed in \cite{Gauntlett:2009bh}, where
a characteristic superfluid dome was seen. Outside of the dome was a class of charged black holes that
as $T\to 0$ become singular in the IR. We have checked that this singularity actually corresponds to a hyperscaling violating behaviour with $z=1$ and $\theta=-1$.

It will be very interesting to unify the analysis of this paper with that of \cite{Gauntlett:2009bh} and study
the $d=3$ CFTs as a function of $T,\mu, B$ and deformation by ${\cal O}_\sigma$. It will be particularly
interesting to see how the hyperscaling violating behaviours with $z=3/2$ and $z=1$ interpolate between each other.
It will also be very interesting to see if the deformation by ${\cal O}_\sigma$ can drive down the temperature
of the critical metamagnetic point to zero temperature to obtain a metamagnetic 
quantum critical point as observed, for example, in 
$Sr_3Ru_2O_7$ \cite{stronruth,PhysRevLett.88.217204} and studied in a different
in a different holographic context in \cite{D'Hoker:2010rz}, building on \cite{D'Hoker:2009mm,D'Hoker:2009bc}.

\subsection*{Acknowledgements}
We would like to thank Alejandra Castro for interesting conversations.
AD is supported by an EPSRC Postdoctoral Fellowship, JPG is supported by a Royal Society Wolfson Award
and BW is supported by a Royal Commission for the Exhibition
of 1851 Science Research Fellowship.
This work was supported in part by STFC grant ST/J0003533/1. This work was supported in part by the U.S. Department of Energy (DOE) under cooperative research agreement Contract Number DE-FG02-05ER41360. BW thanks the CTP at MIT for hospitality while this paper was being finished.

\appendix
\section{Dyonic AdS-RN black holes of Einstein-Maxwell theory}
As somewhat of an aside, in this appendix we calculate the magnetisation and magnetic
susceptibility of the canonical dyonic AdS-RN black holes of Einstein-Maxwell theory following
\cite{Hartnoll:2008kx,Denef:2009yy}. In particular, we want to contrast this with our results 
for the dyonic black holes in the top-down model \eqref{action}, \eqref{fnsub}. 
To obtain Einstein-Maxwell theory, in \eqref{action} we set $\sigma=0$ and 
\begin{align}\label{fnsubem}
V  \equiv   -24\,,\qquad
\tau \equiv 1\,,\qquad
\vartheta  \equiv 0\,.
\end{align}
The dyonic AdS-RN black hole solution is given by
\begin{align}\label{dyonicadsrn}
g=4r^2-(4r_+^2+\frac{\mu^2}{4}+\frac{B^2}{4r_+^2})\frac{r_+}{r}+
(\frac{\mu^2}{4}+\frac{B^2}{4r_+^2})\frac{r_+^2}{ r^2},\qquad \phi=\mu(1-\frac{r_+}{r})\,.
\end{align}
The temperature is given by
\begin{align}
T=\frac{48r_+^4-\mu^2r_+^2-B^2}{16\pi r_+^3}\,,
\end{align}
and this provides us with an expression $r_+=r_+(T,\mu,B)$.
%
Calculating the free-energy $w=w(T,\mu,B)$ as in section \ref{thermsec} we obtain
\begin{align}
w=-4r_+^3-\frac{\mu^2r_+}{4}+\frac{3B^2}{4r_+}\,,
\end{align}
and for the magnetisation $m$ and magnetic susceptibility $\chi_m$ at constant $T,\mu$ we get
\begin{align}
m&=-\left(\frac{\partial w}{\partial B}\right)_{T,\mu}=-\frac{B}{r_+}\,,\nn
\chi_m&=\left(\frac{\partial m}{\partial B}\right)_{T,\mu}=-\frac{48 r_+^4+\mu^2r_+^2+{B^2}}{r_+(48r_+^4+\mu^2r_+^2+3B^2)}\,.
\end{align}
Notice that since $m$ is always negative the system is strongly diamagnetic (as noted in \cite{Hartnoll:2008kx}).
Also, $\chi_m$ is always negative asymptoting to zero for large $B,T$.

We can also use the ensemble $f=f(T,q,B)\equiv w+\mu q$. We find
\begin{align}
f=-4r_+^3+\frac{3(q^2+B^2)}{4r_+}\,,
\end{align}
and
\begin{align}
m&=-\left(\frac{\partial f}{\partial B}\right)_{T,q}=-\frac{B}{r_+}\,,\nn
\chi'_m&=\left(\frac{\partial m}{\partial B}\right)_{T,q}=-\frac{48 r_+^4+3q^2+{B^2}}{r_+(48r_+^4+3(q^2+B^2))}\,,
\end{align}
and we note that $\chi'_m$ has similar behaviour to $\chi_m$, as one expects.

\bibliographystyle{utphys}
\bibliography{helical}{}
\end{document}